\documentclass[longauth]{aa}  

\usepackage{graphicx}
\usepackage{txfonts}
\usepackage{gensymb}
\usepackage[]{hyperref}
\usepackage{multirow}
\usepackage{graphicx}
\usepackage{times}
\usepackage{float}
\usepackage{pdflscape}
\usepackage{threeparttable}
\usepackage{amsmath}
\usepackage{amssymb}
\usepackage{xspace}
\usepackage{caption}
\usepackage{subcaption}
\usepackage{newtxtext}
\usepackage{color}
\usepackage{lscape}
\usepackage{tabularx}

		\newcommand{\Msun}{\mbox{$M_{\odot}$}\xspace}
            \newcommand{\Lsun}{\mbox{$L_{\odot}$}\xspace}

		\newcommand{\fluxunit}{\mbox{\,erg\,cm$^{-2}$\,s$^{-1}$}\xspace}

		\newcommand{\ka}{K$\alpha$\xspace}

		\newcommand{\feka}{Fe\,K$\alpha$\xspace}

		\newcommand{\pmc}{P$_{ \mathcal{MC}}$\xspace}
		\newcommand{\pft}{P$_{ \mathcal{F}}$\xspace}  
		
		\newcommand{\mc}{$\mathcal{MC}$\xspace}

		\newcommand{\vout}{v_{\rm out}\xspace}
		\newcommand{\mout}{\dot M_{\rm out}}
  		\newcommand{\mw}{\mathcal{\dot M}_{\rm w}}			
		\newcommand{\vturb}{v_{\rm turb}\xspace}
		\newcommand{\ecut}{E_{\rm cut}}

		\newcommand{\chisq}{\chi^{2}}
		\newcommand{\nhgal}{N_{\rm H}^{\rm Gal}}
		\newcommand{\nh}{N_{\rm H}}
		\newcommand{\lognh}{\log(N_{\rm H}/\rm{cm}^{-2})}
		\newcommand{\logxi}{\log(\xi/\rm{erg\,cm\,s}^{-1})}
		
		\newcommand{\hea}{He$\alpha$\xspace}
		\newcommand{\lya}{Ly$\alpha$\xspace}
  		\newcommand{\heb}{He$\beta$\xspace}
		\newcommand{\lyb}{Ly$\beta$\xspace}

		\newcommand{\lbol}{L_{\rm bol}} 
            \newcommand{\kbol}{k_{\rm bol}}
		 
            \newcommand{\ekin}{\dot E_{\rm kin}}

		\newcommand{\eddratio}{\lambda_{\rm Edd}}
		
		\newcommand{\lion}{L_{\rm ion}}
		\newcommand{\ledd}{L_{\rm Edd}}
		\newcommand{\medd}{\dot M_{\rm Edd}}
		\newcommand{\mbh}{M_{\rm BH}}
		\newcommand{\pout}{\dot P_{\rm out}}

		\newcommand{\vesc}{v_{\rm esc}}
		\newcommand{\vinf}{v_{\infty}}

		\newcommand{\macc}{\dot M_{\rm acc}\xspace}

  		\newcommand{\blue}{\textcolor{blue}}

		\newcommand{\dc}{$\Delta \mathcal{C}$\xspace}
		\newcommand{\cstat}{\mathcal{C-}\rm stat\xspace}
		
		\newcommand{\sn}{SNR\xspace}

   		\newcommand{\fv}{f_{\rm v}}
  		\newcommand{\lx}{\mathcal{L}_{\rm X}}
		\newcommand{\los}{line-of-sight\xspace}

            \newcommand{\lir}{L_{\rm 8-1000\mu m}}
            \newcommand{\ufos}{UFO$_{\rm S}$\xspace}
		  \newcommand{\ufoh}{UFO$_{\rm H}$\xspace}
		  \newcommand{\ufol}{UFO$_{\rm L}$\xspace}
            \newcommand{\alphaox}{$\alpha_{\rm OX}$\xspace}

		\newcommand{\ergs}{\,\rm erg\,s^{-1}\xspace}
		\newcommand{\cmsq}{{\,\rm cm^{-2}}\xspace}
		\newcommand{\cmq}{{\,\rm cm^{-3}}\xspace}
		\newcommand{\flux}{{\,\rm erg\,cm^{-2}\,s^{-1}}\xspace}
		\newcommand{\rg}{\,R_{\rm g}}
		\newcommand{\ev}{\,\rm eV}		
		\newcommand{\kev}{\,\rm keV}
				
		\newcommand{\kms}{\,\rm km\,s^{-1}\xspace}

		\newcommand{\kb}{$\mathcal{KB}$\xspace}

		\newcommand{\suzaku}{\emph{Suzaku}\xspace} 
		\newcommand{\nustar}{\emph{NuSTAR}\xspace}		
		\newcommand{\xmm}{\emph{XMM-Newton}\xspace}

		\newcommand{\xrade}{\textsc{xrade}\xspace}
  		\newcommand{\wine}{\textsc{wine}\xspace}

		\newcommand{\pexrav}{\texttt{pexrav}\xspace}
  		\newcommand{\phabs}{\texttt{phabs}\xspace}
  		\newcommand{\constant}{\texttt{constant}\xspace}

		\newcommand{\xillver}{\texttt{xillver}\xspace}
		\newcommand{\relxill}{\texttt{relxill}\xspace}

            \newcommand{\dwind}{\textsc{disk-wind}\xspace}		
            
		\newcommand{\dwf}{\texttt{fast32}\xspace}

            \newcommand{\cloudy}{\textsc{Cloudy}\xspace}

		\newcommand{\xstar}{\textsc{xstar}\xspace}
		\newcommand{\xspec}{\textsc{xspec}\xspace}

		\newcommand{\sas}{\textsc{sas}\xspace}

		\newcommand{\omi}{\textsc{omichain}\xspace}

        \newcommand{\sw}{$\sigma_{\rm width}$\xspace}

\hypersetup{
    colorlinks=true,
    linkcolor=blue,
    filecolor=magenta,      
    urlcolor=cyan,
    citecolor=blue
}

\defcitealias{Tombesi15}{T15}

\begin{document} 

   \title{The \xmm and \nustar view of IRASF11119+3257. I\\
   Detection of multiple UFO components and a very cold corona.}

\titlerunning{The UFO in IRASF11119}
\authorrunning{G. Lanzuisi et al.,}

   \author{G.~Lanzuisi\inst{1}
          G.~Matzeu\inst{1,2,3},
          P.~Baldini\inst{1,2,4},
          E.~Bertola\inst{5},
          A.~Comastri\inst{1},
          F.~Tombesi\inst{6,7,8,9,10},
          A.~Luminari\inst{11,7},
          V.~Braito\inst{12,13,14},
          J.~Reeves\inst{12,13},
          G.~Chartas\inst{15},
          S.~Bianchi\inst{16},
          M.~Brusa\inst{2,1},
          G. Cresci\inst{5}
          E.~Nardini\inst{5}
          E.~Piconcelli\inst{7},
          L.~Zappacosta\inst{7},
          R.~Serafinelli\inst{7},
          M.~Gaspari\inst{17},
          R. Gilli\inst{1},
          M.~Cappi\inst{1},
          M.~Dadina\inst{1},
          M.~Perna\inst{18},
          C.~Vignali\inst{2,1} and
          S.~Veilleux\inst{7,19}
          }

\institute{INAF -- Osservatorio di Astrofisica e Scienza dello Spazio di Bologna, Via Gobetti, 93/3, I-40129 Bologna, Italy, \email{giorgio.lanzuisi@inaf.}
\and
Department of Physics and Astronomy (DIFA), University of Bologna, Via Gobetti, 93/2, I-40129 Bologna, Italy
\and
Quasar Science Resources SL for ESA, European Space Astronomy Centre (ESAC), Science Operations Department, 28692, Villanueva de la Ca\~{n}ada, Madrid, Spain
\and
Max Planck Institute f\"ur Extraterrestriche Physik, Giessenbachstrasse, D-85748, Garching, Germany
\and
INAF -- Osservatorio Astrofisico di Arcetri, Largo Enrico Fermi 5, 50125 Firenze, Italy
\and
Department of Physics, University of Rome ‘Tor Vergata’, Via della Ricerca Scientifica 1, I-00133 Rome, Italy
\and
INAF -- Osservatorio Astronomico di Roma, Via Frascati 33, 00078, Monte Porzio Catone (Roma), Italy
\and
INFN - Rome Tor Vergata, Via della Ricerca Scientifica 1, 00133 Rome, Italy 
\and
Department of Astronomy, University of Maryland, College Park, MD 20742, USA
\and
NASA Goddard Space Flight Center, Code 662, 
Greenbelt, MD 20771, USA
\and
INAF -- Istituto di Astrofisica e Planetologia Spaziali, Via Fosso del Cavaliere, I-00133 Roma, Italy
\and
Department of Physics, Institute for Astrophysics and Computational Sciences, The Catholic University of America, Washington, DC 20064, USA
\and
INAF -- Osservatorio Astronomico di Brera, Via Bianchi 46, I-23807 Merate, LC, Italy
\and
Dipartimento di Fisica, Universitá di Trento, Via Sommarive 14, Trento 38123, Italy.
\and
Department of Physics and Astronomy, College of Charleston, Charleston, SC, 29424, USA
\and
Dipartimento di Matematica e Fisica, Universit\`{a} degli Studi Roma Tre, Via della Vasca Navale 84, I-00146, Roma, Italy
\and
Department of Astrophysical Sciences, Princeton University, 4 Ivy Lane, Princeton, NJ 08544-1001, USA
\and
Centro de Astrobiologia (CAB), CSIC–INTA, Departamento de
Astrofisica, Cra. de Ajalvir Km. 4, 28850 – Torrejon de Ardoz, Madrid,
Spain
\and
Joint Space-Science Institute, University of Maryland, College Park, MD 20742, USA
             }

   \date{Received Jan. 9th 2024; accepted Jun. 17th 2024}

\abstract
{IRASF11119+3257 is an ultra-luminous infrared galaxy with post-merger morphology, hosting a type-1 QSO at z=0.189. It shows a prominent Ultra Fast Outflow (UFO) absorption feature ($\vout\sim0.25c$) in its 2013 \suzaku spectrum. This is the first system in which the energy released by the UFO was compared to that of the known galaxy-scale molecular outflow to investigate the mechanism driving AGN feedback.}
{In 2021, we obtained the first \xmm long look of the target, coordinated with a simultaneous \nustar observation, with the goal of constraining the broad-band continuum and the nuclear wind physical properties and energetics with unprecedented accuracy.}
{The new high-quality data allow us to clearly detect at \pmc$>99.8\%$ confidence level multiple absorption features associated with the known UFO at 9.1 and 11.0 keV rest-frame. Furthermore, an emission plus absorption feature at $1.1-1.3$ keV reveals the presence of a blueshifted P-Cygni profile in the soft band.}  
{We associate the two hard band features with blends of FeXXV and FeXXVI \hea-\lya and \heb-\lyb line pairs and infer a large column ($\nh\sim10^{24}$ $\cmsq$) of highly ionized ($log\xi\sim5$) gas outflowing at $\vout=0.27\pm0.01c$. The 1.3 keV absorption line can be associated with a blend of Fe and Ne transitions, produced by a lower column ($\nh\sim3\times10^{21}$ $\cmsq$) and ionization ($log\xi\sim2.6$) gas component outflowing at the same speed. Using a radiative-transfer disk wind model to fit the highly ionized UFO, we derive a mass outflow rate comparable with the mass accretion rate and the Eddington limit ($\mout=4.25_{-0.73}^{+1.11} \Msun/yr$, $\sim1.6$ $\macc$, $\sim1.0$ $\medd$), and kinetic energy ($\ekin=1.21_{-0.20}^{+0.32}\lbol, \sim0.7 \ledd$)  and momentum flux ($\pout=6.37_{-1.09}^{+1.67} \lbol/c$) among the highest reported in the literature. We measure an extremely low high-energy cut-off ($\ecut\sim25-30$ keV). This and several other cases in the literature suggest that a steep X-ray continuum may be related to the formation of powerful winds. We also analyzed the ionized [OIII] component of the large-scale outflow through optical spectroscopy and derived a large outflow velocity ($\vout\sim3000$ km/s) and energetics comparable with the large-scale molecular outflows. Finally, we observe a trend of decreasing outflow velocity from forbidden optical emission lines of decreasing ionization levels, interpreted as the outflow decelerating at large distances from the ionizing source.}
{The lack of a significant momentum boost between the nuclear UFO and the different phases of the large-scale outflow, observed in IRASF11119 and in a growing number of similar sources, can be explained by (i) a momentum-driven expansion, (ii) an inefficient coupling of the UFO with the host interstellar medium, or (iii) by repeated energy-driven expansion episodes with a low duty-cycle, that average out on long time-scales to produce the observed large-scale outflow.}


   \maketitle

\section{Introduction}

Ultra Fast Outflows (UFOs, $\vout>10^4$\, km\,s$^{-1}$) launched in the vicinity of accreting Super Massive Black Holes (SMBH) were first discovered in the luminous lensed quasars APM 08279+5255 \citep{Hasinger02, Chartas02} and in two low-redshift analogs, PG 1211+143 \citep{Pounds03} and PDS 456 \citep{Reeves03}. They are now routinely detected as blue-shifted iron K-shell absorption lines at energies above $\sim7$ keV in the X-ray spectra of 30–40\% of low redshift (z<0.5) Seyferts \citep{Tombesi10, Gofford13}, quasars \citep{Matzeu23}, and radio Galaxies \citep{Tombesi14MNRAS.443.2154T} and in an even larger fraction of the few - mostly lensed - high redshift (i.e. $z\gtrsim2$) quasars with good enough X-ray spectra \citep{Bertola20, Chartas21}. 

These nuclear winds span a wide range in velocity (from the defining threshold of $\vout>10^4$ $\kms$, up to $\sim0.3-0.5c$, e.g., \citealt{Reeves18PDS, Chartas09}), with gas column densities and ionization parameters in the ranges $\lognh\sim23-24$ and $\logxi\sim3-6$, respectively. Given their physical properties and their likely wide-angle geometry nature - consistently derived from both the detection rate in the general AGN population and the observed P-Cygni profile shape in a few well-studied examples \citep[e.g.,][]{Nardini15, Laurenti21} - they are expected to deposit a significant fraction of the SMBH accretion power into the host galaxy \citep[e.g.,][]{Gaspari12}. 

Warm Absorbers (WA) with much lower column densities ($\lognh\sim20-22$) and ionization parameters ($\logxi\sim 1-3$) are observed as absorption features and edges from He- and H-like ions of 
C, O, N, Ne, Mg, and other elements mostly in the soft X-ray band \citep[e.g.,][]{Blustin05}. 
These outflows have much lower velocity (up to $5\times10^3$ km/s), and even if they are detected in an even larger fraction of AGN, i.e. $\sim65\%$ \citep{Reynold97, Piconcelli05, McKernan07}, they are estimated to carry only a small fraction of the kinetic energy of the outflow. 

The claim of high-velocity ionized outflow detected in the soft band dates back to the very beginning of the nuclear winds studies era \citep{Pounds03} and has been recently reported in a growing number of local Seyferts and quasars
\citep{Gupta13b, Gupta15, Longinotti15, Reeves16, Pounds16, Parker17PCAIRAS13224, Reeves18PG1211, Serafinelli19, Krongold21}.

These absorbers, often observed in high-resolution grating spectra in the soft X-ray band, are attributed to gas clouds with much lower ionization and column densities with respect to UFOs but with comparable outflow velocities. 
This class of outflows can be interpreted as being due to dense clumps of shocked interstellar medium entrained at large distances (hundreds of pc) by the advancing outer shock \citep{Fauchere12, Zubova12}, or as the imprint of the UFO gas itself cooling down and fragmenting at hundreds of $\rg$ from the SMBH \citep{Takeuchi13, Kobayashi18, Gaspari17_uni}. In both cases, Rayleigh-Taylor and Kelvin-Helmholtz's instabilities are central to the process, as well as turbulence and radiative multiphase condensation (e.g., via chaotic cold accretion (CCA) mechanism; \citealt{Gaspari17_cca}).

All these multi-phase and multi-scale winds driven by the central AGN are thought to play a fundamental role in shaping the SMBH/galaxy co-evolution \citep[e.g.,][]{Fabian12, Gaspari20, Veilleux20, Laha21NatAs} by removing and/or heating the cold gas from the host, quenching the growth of both the SMBH and the stellar component - the so-called AGN feedback -  and possibly
driving the scaling relations between SMBH mass and host properties \citep[e.g.,][]{Silk98, Magorrian98}.

Finding observational evidence for AGN feedback in action is of crucial importance to understanding galaxy and SMBH evolution \citep[e.g.,][]{Fiore17}.
Theoretical models suggest energy- or momentum-conserving flows powered by the nuclear UFO \citep[e.g.,][]{Zubova12, Costa14} as the driving mechanism of the large-scale outflows. Momentum-conserving flows occur if most of the wind kinetic energy is radiated away. Energy-conserving flows occur if the gas shock-heated by the wind is not efficiently cooled and instead expands adiabatically as a hot bubble.  In the latter case, the momentum rate of the large-scale outflow is boosted by a factor $\propto v_{\rm dw}/v_{\rm ls}$, where $v_{\rm dw}$ is the velocity of the nuclear disk wind and $v_{\rm ls}$ the velocity of the large scale outflow.

Originally, studies of large-scale molecular outflows lacked the detection of the inner winds \citep[e.g.,][]{Veilleux2013}. Conversely, studies of AGN accretion disk winds focused only on X-ray observations of local AGNs \citep[e.g.,][]{Tombesi10, Gofford13} and a few high redshift quasars \citep{Chartas02, Lanzuisi12, Bertola20}. This situation changed when the detection of an accretion disk wind and a powerful molecular outflow was reported in the same source, IRASF11119+3257 (IRASF11119 hereafter; \citealt[hereafter \citetalias{Tombesi15}]{Tombesi15}), finding momentum flux values supporting the energy-driven scenario. Later studies \citep{Veilleux17} revised the energetics of the large-scale outflow - also seen in CO(1-0) emission line wings with ALMA - by time-averaging over the flow time and adopted different assumptions on the covering factor and coupling of the UFO with the ISM, 
obtaining energetics for the wind more in line with the momentum-driven scenario (see also \citealt{NardiniZubovas} who also revised the BH mass).

These results started an entirely new line of research trying to observationally link the small-scale relativistic winds and the large-scale outflows in a few bright local and high-z AGNs \citep[e.g.,][]{Feruglio17, Chartas20} through the use of the `Momentum boost' diagram and similar metrics (e.g., the energy-transfer rate). 
The observational evidence, however, has shown contradictory results, with different systems showing a variety of behaviors in terms of outflow energetics \citep[e.g.,][]{ReevesBraito19, Smith19, Bischetti19PDS, Marasco20, Tozzi21}.
Because of the large uncertainties affecting these measurements and the limited samples (see, e.g., \citealt{bonanomi23} for a recent compilation)
up to now, the results are inconclusive.
Variability on different time scales is an unavoidable complication in these studies, as both the nuclear wind power and the SMBH accretion rate vary on time scales that are much shorter than the galaxy-scale winds flow time \citep[see, e.g.,][]{ZubovasNardini}. 

In this paper, we analyze new \xmm and \nustar data 
of IRASF11119 to derive accurate physical parameters and energetics for the persistent nuclear outflow observed in X-rays and test the energy-driving scenario for the impact of such outflow on large scales. The paper is organized as follows: Sec.~\ref{sec:IRASF} illustrates the target properties; Sec.~\ref{sec:data_red} describes the procedures adopted in data reduction; in Sec.~\ref{sec:spectral} we describe the continuum modeling and the search for absorption and emission features; in  Sec.~\ref{sec:line_sign} we derive the main properties and statistical significance of the detected features, while in Sec.~\ref{sec:UFOmodel} we model the UFO features with physical wind models;
in Sec.~\ref{sec:coronal}, we discuss the possible connection between outflows and coronal properties; in Sec.~\ref{sec:ionized}, we derive the properties of the large-scale ionized outflow from optical spectroscopy; in Sec.~\ref{sec:ene}, we derive the energetics of the nuclear and large-scale ionized outflows and compare them with results from other targets. and finally, in Sec.~\ref{sec:conc}, we summarize our findings, 
We adopt the cosmological parameters $H_0 = 70$ km s$^{-1}$ Mpc$^{-1}$, $\Omega_{\Lambda} = 0.73$ 
and $\Omega_m = 0.27$. 
Errors are given at a 90\% confidence level. 

\section{IRASF11119+3257}
\label{sec:IRASF}
IRASF11119 (coordinates RA=11:14:38.9, Dec=+32:41:33.3)
is a nearby ($z=0.189$) Ultra-luminous Infrared Galaxy (ULIRG, $\lir >10^{12} \Lsun$), with post-merger morphology, hosting a red type-1 AGN with quasar luminosity ($\lbol=1.5\times10^{46}$ erg/s), responsible for $\sim80\%$ of the total emission of the system \citep{Veilleux2013}. 
It has a full width at half-maximum (FWHM) of the $H\beta$ line of $\sim2000 \kms$, quite uncertain due to the complex emission lines structure (see Sec. \ref{sec:ionized}). This is right at the defining threshold for Narrow Line Seyfert 1s (NLSy1s), and indeed, IRASF11119 was originally classified as such \citep{Zheng02}. The most updated BH mass estimate from single epoch spectroscopy is $\mbh=1.95_{-0.54}^{+0.80}\times10^{8}$ \Msun \citep{NardiniZubovas} which, for the source luminosity, translates into an Eddington ratio of $\eddratio\sim0.6$.

A powerful UFO with mildly relativistic velocity ($\vout\sim0.25c$) was discovered through the detection of a strong absorption feature at $\sim9$ keV rest-frame, in a 250 ks \suzaku spectrum taken in 2013 \citepalias{Tombesi15}. A massive molecular outflow was previously observed in absorption in the OH 119 $\mu m$ transition with Herschel with velocity $\vout(OH)=1000\pm200$ km s$^{-1}$ \citep{Veilleux2013}. 
From these early estimates, the momentum rates derived for the accretion disk wind and molecular outflow appeared consistent with an energy-conserving scenario, with the large-scale outflow momentum boosted by a factor $\propto$ $v_{\rm dw}/v_{\rm ls}$ with respect to the inner flow. This was the first direct evidence of a UFO possibly driving a large-scale molecular outflow in its host galaxy through an energy-driven expansion \citepalias{Tombesi15}. The UFO was later confirmed with \nustar\ data \citep{Tombesi17} taken in 2016. Due to the lower energy resolution of \nustar ($\sim0.4$ keV at 6 keV),
the UFO was seen as a broad blended feature of absorption lines and edges above $\sim9$ keV. The properties of the nuclear outflow ($\vout$, log$\xi$, $\ekin$, and $\pout$) are broadly consistent between the two X-ray measurements, with significant variation observed in the column density, going from $\sim3$ to $\sim6$ $\times 10^{24} \cmsq$, suggestive of a clumpy outflowing gas. 

The source also shows intense ionized outflows seen as broad emission line wings of [OIII] ($\vout=1300$ km s$^{-1}$, \citealp{Lipari03}) and Na~I~D absorption features ($\vout~\leq~1450$ km s$^{-1}$ \citealp{Rupke05}) in the optical, and [Ne III] and [Ne V] emission line wings ($\vout=679$ km s$^{-1}$ and $=870$ km s$^{-1}$ respectively, \citealp{Spoon09}) in the Mid-IR. 

An independent measurement of the molecular outflow energetics was obtained by detecting CO(1-0) broad wings in ALMA data \citep{Veilleux17}. The CO outflow has the same outflow velocity $\vout\sim1000$ km s$^{-1}$ but a factor of $\sim5$ smaller momentum rate with respect to OH. It is possible to reconcile the two measurements, however, once the momentum of the OH outflow is time-averaged, taking into account the flow timescale $R_{out}/V_{out}$. The estimated extents of the OH and CO outflows are 0.1-1 kpc and $\sim7$ kpc, respectively.

Adopting the revised BH mass mentioned above and different assumptions on the covering factor and coupling with the ISM, \citet{NardiniZubovas} derived a higher momentum rate for the X-ray UFO. This new estimate, together with the time-averaged OH and the CO outflow values, make the energetics of this wind inconsistent with the energy-driven scenario at face value. These authors argue that if energy-driven expansion is to be expected beyond a certain radius - (tens/hundreds pc, e.g., \citealt{Fauchere12,Costa14}), 
any major deviations from the expected trend could be attributed to fluctuations in the past AGN activity, suggesting that the nuclear outflow observed today may be significantly stronger than it was on average in the past and that the large-scale outflow is the result of several distinct, short-lived high-activity phases. 

These results show the potential of this approach for the understanding of the galactic wind driving mechanism(s) and for reconstructing the history of SMBH accretion, but also highlight the importance of having excellent quality data to shrink the statistical error bars, keeping in mind the systematics related to the many assumptions involved in the derivation of the winds energetics (see \citealt{Harrison18} for a review). 

More recently, \citet{pan2019discovery} reported
the discovery of a HeI $10830\AA$ mini-BAL thanks to near-IR spectroscopy, with velocities and extensions similar to those measured for the emission line outflow (1000 km/s, $\sim3$ kpc).
Finally, thanks to very-long-baseline interferometric (VLBI) observations with the European VLBI Network (EVN) at 1.66 and 4.93 GHz, \citet{Yang20} recently reported the detection of a two-sided but significantly beamed jet. The authors infer that the approaching jet has an intrinsic speed $\vout>0.57c$. The source has a quite high radio power of log(L(1.4GHz))=25.04 W Hz$^{-1}$ from NVSS \citep{Condon98} and is therefore classified as moderately radio-loud ($R_{1.4}\sim20$) once the obscuration in the optical bands is taken into account \citep{Komossa06}.

\section{New Observations and Data reduction}
\label{sec:data_red}

Observing  IRASF11119 with a long \xmm\ observation, coordinated with \nustar, was part of a systematic approach devoted to searching and characterizing UFOs in QSOs beyond the local Universe. This led to the approval of the SUBWAYS program \citep{Matzeu23,Mehdipour22,Gianolli23}, a 1.8 Ms \xmm\ large program approved in AO18 to observe with unprecedented spectral quality 20 QSOs at $z=0.1-0.5$. IRASF11119 was the only bright source in the same redshift and luminosity range missing a dedicated \xmm\ long observation and, therefore it was requested in AO20 to enrich the sample with this peculiar and well-studied target.

IRASF11119 was observed by \xmm\ (obs. ID 0881350101) on 2021-11-12 for 117ks and by \nustar\ (obs. ID 60701006002) on the same date for 53ks (PI G. Lanzuisi). \xmm pn \citep{Struder01} and MOS1-2 \citep{Turner2001} data were reduced using the standard software \sas, v20.0.0.
For high-background screening, we followed the \citet{Piconcelli04} procedure to maximize the source \sn by iteratively finding the combination of source extraction region and background filtering that maximizes the \sn in a given band.

We decided to perform this optimization in the $4-10\kev$ band in all cameras and not generically in the full $0.5-10\kev$ band as the former is the band where the UFO search will be carried out, given the source redshift. At these energies, the Effective Area is smaller and the background contribution larger, and therefore the optimization of the \sn substantially differs from the one obtained for the full band
(see also \citealt{Matzeu23}). 
We selected only events corresponding to single and double-pixel events (pattern 0–4 and 0–12 for pn and MOS, respectively). A background region of $60$ arcsec radius, free of instrumental features and other detected sources, is adopted. Different extraction radii were tested to define the source extraction region. For each radius, we calculated the maximum level of $4-10\kev$ band background that can be tolerated to find the optimal \sn in the same band. 
The final parameters for pn are: source radius $r_{\rm src}=15''$, (corresponding to 75\% EEF at 1 keV or 67\% at 5 keV), background count rate normalized to the source area $CR_{\rm bkg}=0.05$ c/s, and \sn$=48.3$, while for MOS1(2) the parameters are $r_{\rm src}=16''(23'')$, $CR_{\rm bkg}=0.01(0.003)$ c/s and \sn$=32.6(35.6)$.

The final exposure times are 92ks for the pn and 113(111) ks for MOS1(2), respectively. Given the source flux ($F_{4-10keV}=6\times10^{-13}\flux$), our approach allowed us to retain most of the exposure time ($78$ and $96\%$ respectively), despite the presence of several background flares during the observation. A standard background flare cut, on the other hand, would have retained only
$\sim50$\% and $\sim70$\% of the exposure time, respectively.

The EPIC source spectra were inspected for the presence of photon pile-up using the \sas task \textsc{epatplot}. The observed-to-model singles and doubles pattern are both consistent with 1.0 within statistical errors and thus no significant pile-up is present.
The response files were subsequently generated with the \sas tasks \textsc{rmfgen} and \textsc{arfgen} with the calibration EPIC files version v3.13.

The NuSTAR data were processed using the NuSTAR Data Analysis Software package (NuSTARDAS) version 2.1.2 within Heasoft v.6.30 tools\footnote{\url{https://heasarc.nasa.gov/lheasoft/}}. Calibrated and cleaned event files were produced using the calibration files in the NuSTAR CALDB (version 20220510) and standard filtering criteria with the \textsc{Nupipeline} task. We checked for high background periods using the \textsc{nustar\_filter\_lightcurve} IDL script\footnote{\url{https://github.com/NuSTAR}}. No significant solar flares were detected; the final NuSTAR exposure time is 52ks. 
We tested different extraction regions to optimize the source spectral \sn at high energies. Finally, we adopted a region of 60 arcsec radius, corresponding to $\sim70\%$ of the encircled energy fraction for the NuSTAR point spread function \citep{An14}. This allowed us to detect the source at over $3\sigma$ up to 20 keV observer-frame. 
The background was created using the \textsc{nuskybgd} IDL script\blue{$^2$}, which simulates the background spectrum at the source position considering the position-dependent stray light background component.

Finally, all the \xmm\ and \nustar\ spectra have been grouped using the KB binning \citep{KaastraBleeker16}, 
which is specifically developed to optimize the \sn in narrow, unresolved spectral features while maintaining the necessary spectral resolution (see Appendix C in \citealt{Matzeu23} for a thorough comparison between this and other binning schemes). 
We performed the fit by using the Cash statistics with
direct background subtraction (also called W-stat, \citealt{Cash79}; \citealt{Wachter79}) in \xspec \citep{Arnaud96}. 
In the following, the unbinned MOS and FPM spectra have been merged and then binned with KB binning for plotting purposes only to highlight emission/absorption features.
We also verified if the RGS spectra contained any useful source information by looking at the pre-reduced data. However, the source is too faint in the soft band ($F_{0.5-2keV}=1.8\times10^{-13}$ \fluxunit), and the extracted spectra are completely background-dominated.

\begin{figure}
\centering
\includegraphics[width=0.48\textwidth]{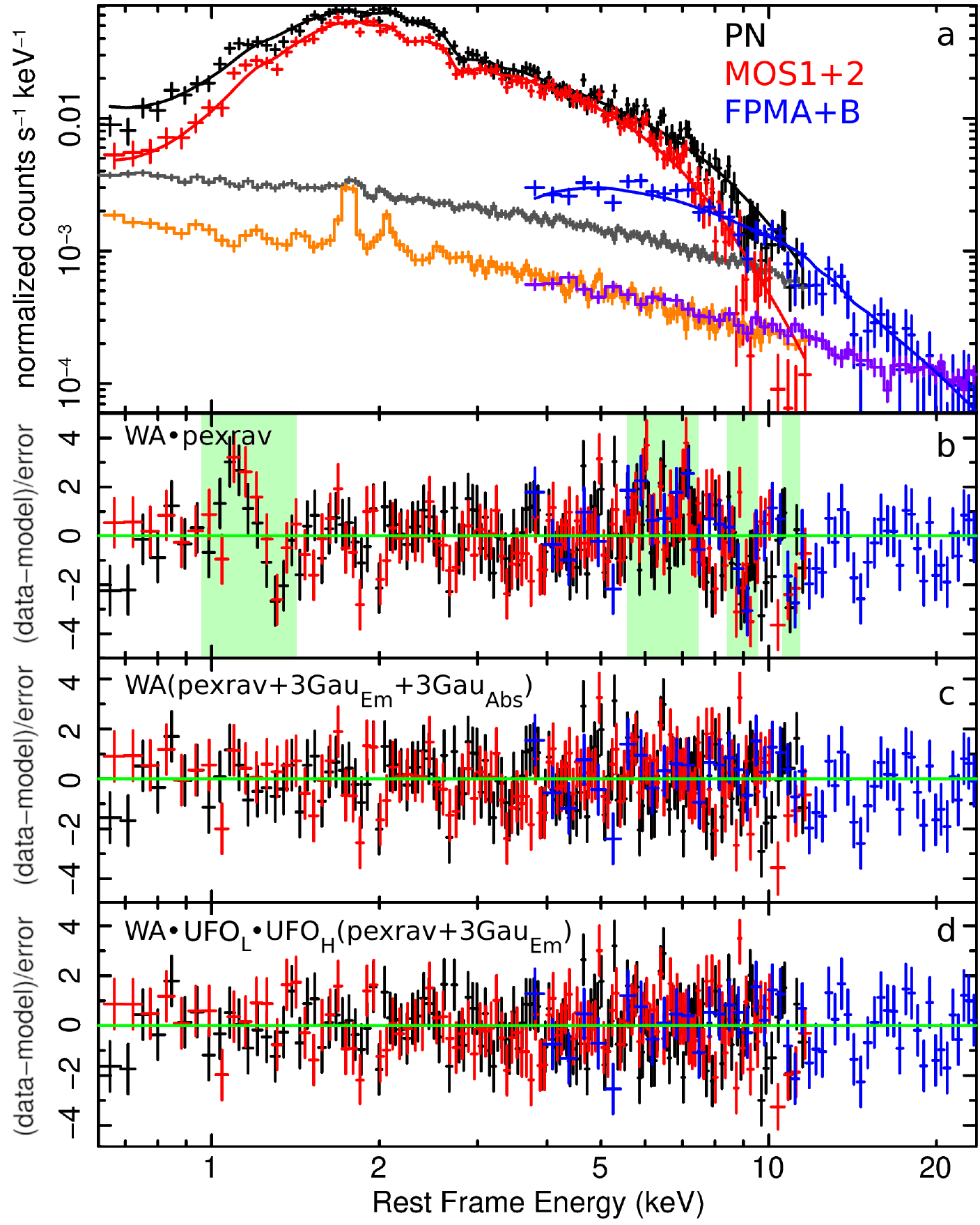}
\caption{{\bf a:} \xmm and \nustar spectra and best-fit continuum model. Background subtracted \xmm (pn in black, MOS1+2 in red) and \nustar (FPMA+B in blue) spectra, and their corresponding background levels (gray, orange, and violet, respectively) are shown. The MOS1 and 2 and FPM-A and -B spectra have been merged for plotting purposes. 
{\bf b:} Residuals (data–model/error), with respect to the baseline continuum model (see Sec.~\ref{subsec:continuum}). In green, we highlight the energy ranges where the most prominent positive and negative residuals are found (see Sec.~\ref{subsec:spec_scan}). 
{\bf c:} Residuals after including three emissions and three absorption Gaussian features (see Sec.~\ref{sec:line_sign}).
{\bf d:} Residuals after including two \xstar tables to model the soft and hard band absorption features (see Sec.~\ref{subsec:xstar}).
}
\label{fig:broadb}
\end{figure}

\begingroup
\renewcommand{\arraystretch}{1.3} 
    
 \begin{table}[t]
 \centering
 \caption{Broad-band 0.6-24 keV rest-frame continuum model best-fit parameters and source properties.}
   \begin{threeparttable}
    \begin{tabular}{llcc}  
        \hline
        \noalign{\smallskip}
        Component & Parameter      &  & Units \\
        \noalign{\smallskip}
        \hline
        \noalign{\smallskip} 
   \constant &   C$_{\rm MOS}$      &  $1.03$      &         -  \\
             &   C$_{\rm Nus}$      &  $1.13$      &         -   \\
    \phabs   &    log($\nh$)        &  $20.3^*$    &  $\cmsq$   \\
    \pexrav  &   $\Gamma$           & $2.05\pm0.05$  &  -  \\ 
            &    R                  & $<0.36$        &  -    \\
            &    $\ecut$            & $30.7_{-5.5}^{+8.3}$  &  \kev \\
            &    $f_{scat}$          &  $1.2\pm0.4\%$  &  -  \\
   WA   &   log($\nh$)           &  $22.53\pm0.05$   &  $\cmsq$  \\ 
             &   log($\xi$)        &  $0.38\pm0.14$     & erg cm/s  \\
            &   $\vturb$          & $300^*$           &  km/s  \\
            &  $\vout$            & $<2500$           &  km/s  \\
        \noalign{\smallskip} 
    \hline
        \noalign{\smallskip} 
            &  $\cstat/\nu$            & $552.5/376$        &  -    \\ 
            &   $F_{\rm 0.5-2\,keV}$      & $1.8\times10^{-13}$ &  \fluxunit \\
            &   $F_{\rm 2-10\,keV}$       & $5.2\times10^{-13}$ &  \fluxunit \\
            &   $L_{\rm 2-10\,keV}$       & $1.2\times10^{44}$  &  $\ergs$  \\   
       \noalign{\smallskip}
       \hline
    \end{tabular}
    \begin{tablenotes}[para,flushleft]
     C$_{\rm MOS}$ and C$_{\rm Nus}$ are cross-calibration constants for MOS and \nustar spectra; log($\nh$) is the logarithm of column density for either the cold absorber \phabs or the Warm Absorber (WA) component; $\Gamma$ is the power-law photon index; R the reflection parameter defined as the solid angle covered by the cold reflecting material in units of $2\pi$; $\ecut$ is the high-energy cut-off; $f_{scat}$ is the relative intensity of the secondary, unabsorbed, \pexrav component; log($\xi$), $\vturb$ and $\vout$ are the logarithm of ionization parameter, turbulent velocity and outflow velocity of the WA, respectively. $\cstat/\nu$ is the fit statistic over the number of degrees of freedom; $F_{\rm 0.5-2\,keV}$ and $F_{\rm 2-10\,keV}$ are the observed soft and hard band fluxes; $L_{\rm 2-10\,keV}$ is the absorption-corrected luminosity in the 2-10 keV rest frame band; $*$~fixed parameters.
    \end{tablenotes}
    \end{threeparttable}
 \label{tab:cont}
 \end{table}

\endgroup

We reduced OM data following standard procedures with \omi. We used the OM UVW1, U, and V data points (the only filters in which the source is clearly detected) together with the de-absorbed X-ray best-fit model (see Sec.~\ref{subsec:continuum}) to construct the source broadband SED and compute the total ionizing luminosity in the range $1-1000$ Ryd, $\lion\sim4\times10^{45} \ergs$, which is one of the input parameters of photo-ionization models adopted later. From the SED we derived 
\alphaox$\sim1.4~$, typical for sources of similar luminosities (where \alphaox is defined as 
\hbox{\alphaox=-$\log(L_{2\rm keV}/L_{2500\mbox{\scriptsize\AA}}$)/$\log(\nu_{2\rm keV}/\nu_{2500\mbox{\scriptsize\AA}})$}, \citealt{tana79}). 
A detailed characterization of the SED, to be used as input for creating dedicated absorption tables with \xstar and other wind models, will be presented in Paper II.

\section{Spectral analysis}
\label{sec:spectral}

\subsection{Continuum}
\label{subsec:continuum}

\begin{figure*}
\includegraphics[height=7.0cm,width=9cm]{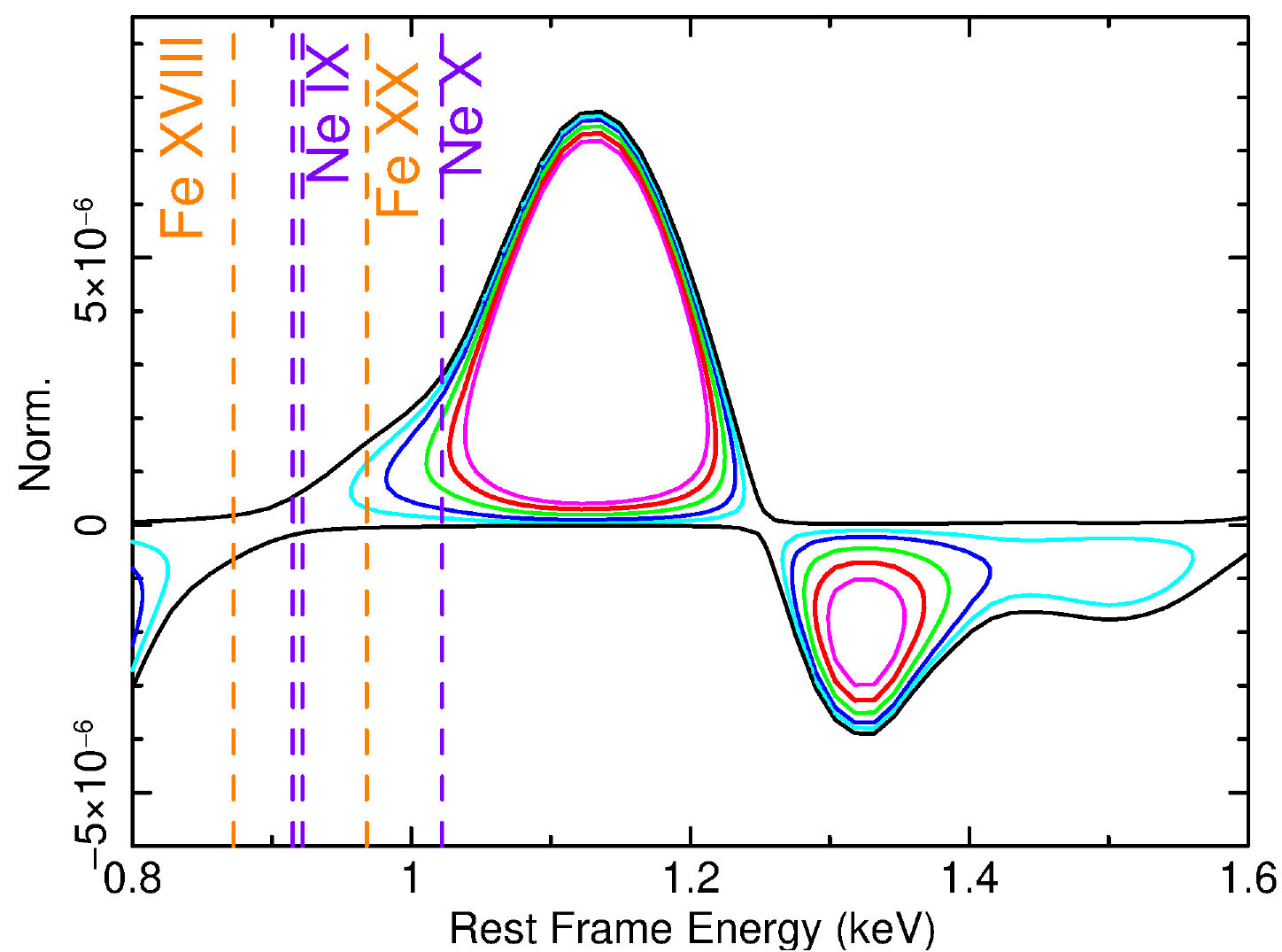}\hspace{0.5cm}
\includegraphics[height=7.0cm,width=9cm]{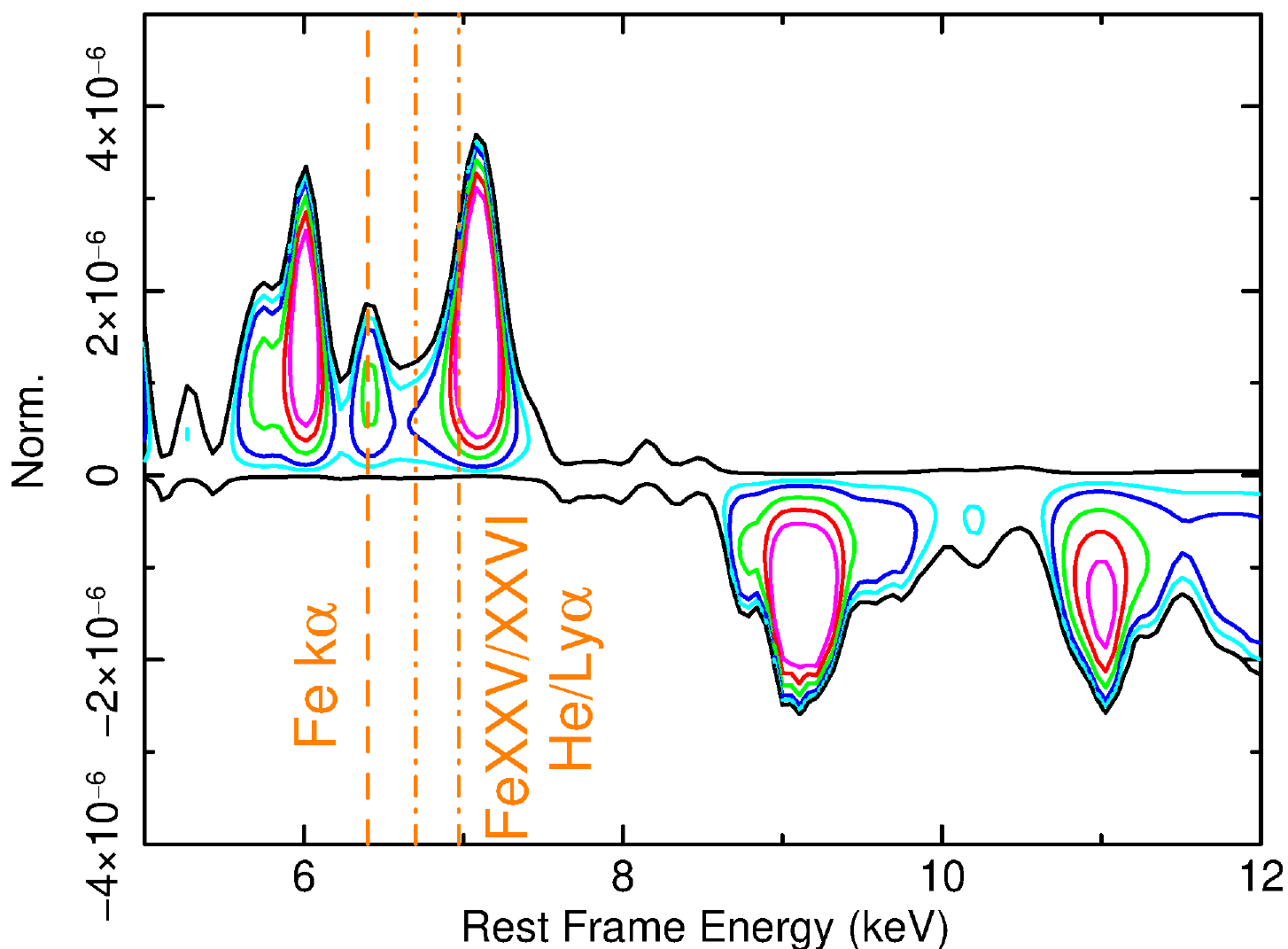}
\caption{Confidence contours showing deviations from the continuum model described in Sec.~\ref{subsec:continuum} in the soft (left) and hard (right) band. Black contours show \dc$=+0.5$ while the colored contours show \dc$=-2.3$ (cyan),$-4.61$ (blue),$-9.21$ (green),$-13.82$ (red),$-18.42$ (magenta), corresponding to a significance of $68\%$, $90\%$, $99\%$, $99.9\%$, and $99.99\%$ for two  parameters. In both panels, the main transitions from ionized Ne and Fe are labeled at their rest-frame energies.
}
\label{fig:spec_scan}
\end{figure*}

The broadband continuum of IRASF11119 (\autoref{fig:broadb}), fitted over the observed 0.5-10 keV band with \xmm and 3-20 keV band with \nustar, can be broadly described by a simple model comprising a primary power-law emission with reflection and high energy cut-off. We model this with \pexrav in \xspec, leaving the 
photon index, reflection parameter, and high energy cut-off as free parameters, while the abundances and inclinations are fixed to the default values (solar abundances and cos(i)=0.45). The best-fit photon index is $\Gamma=2.05\pm0.05$, with a negligible reflection component $R<0.36$, and a very low high-energy cutoff, $\ecut\sim30$ keV. 
The peculiar low value of $\ecut$ will be discussed in more detail in Sec.~\ref{sec:coronal}.
The low reflection fraction is consistent with the very weak neutral Fe \ka emission line (see also \autoref{fig:spec_scan}).
By fixing the reflection parameter R to 1, we obtain an even lower high energy cut-off, $\ecut\sim20$ keV, but we over-predict the flux at 6.4 keV.

This primary emission is modified by a warm absorber (WA) modeled with a \xstar \citep{BautistaKallman01,Kallman04} grid with $\vturb=300 \kms$ and power-law input SED with a photon index of $\Gamma=2$ (\citealt{Matzeu23}).
The best-fit parameters are $\nh=3.4\times10^{22}$ $\cmsq$ and $\logxi=0.4$, with outflow velocity consistent with zero ($\vout<2500$ km/s)\footnote{The velocity is derived from the observed absorber redshift table parameter $z_o$, using the Doppler formula and the relations between intrinsic absorber redshift $z_a$ and source cosmological redshift $z_c$ described e.g. in \citealt{Tombesi11}.}. 
An ionized absorber is preferred with respect to a cold one as the latter is not able to reproduce the observed shape of the soft part of the spectrum: the improvement on the fit statistic ($\Delta\cstat$) when using an ionized absorber instead of a cold one, is $\Delta\cstat=91.3$ for two more free parameters: ionization state and outflow velocity. A secondary, unobscured \pexrav component, with the same parameters and normalization equal to $\sim1.2\%$ of the primary, is needed to model the rise below 1keV and interpreted as scattered emission. This is equivalent to a partial covering absorber with a covering factor $CF=0.99$. 

The continuum model can be described in the following way in \xspec format:

\begin{equation}
\hspace{0.3cm}\texttt{constant}\times\texttt{phabs}\times(\texttt{pexrav}+(\texttt{xstar$_{\rm WA}$}\times\texttt{pexrav})),
\label{eq:baseline}
\end{equation}\vspace{-0.3cm}

\noindent where the \texttt{constant} is introduced to take into account cross-calibration between the EPIC cameras ( within 3\%) and between the \xmm and \nustar observation (within 13\%). The \texttt{phabs} component models the cold galactic absorption: all the fits presented in this section include the Galactic column density of $\nhgal=2\times10^{20} \cmsq$ obtained from the \citet{HI4PI16} survey.  
The best-fit parameters and continuum properties, such as flux and luminosity, are reported in \autoref{tab:cont}.

\begin{figure*}
\includegraphics[height=7cm,width=9cm]{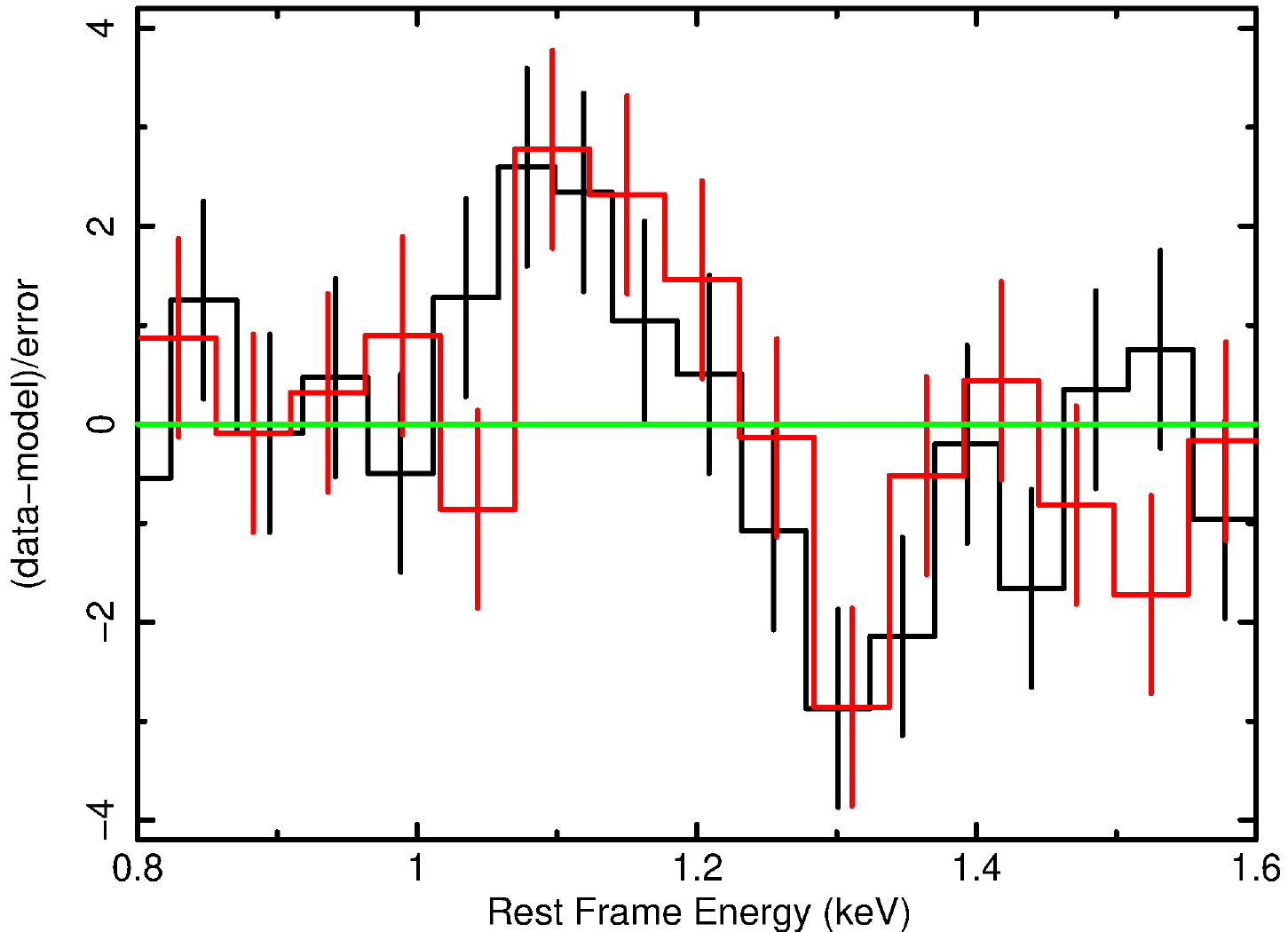}\hspace{0.3cm}
\includegraphics[height=7.cm,width=9.2cm]{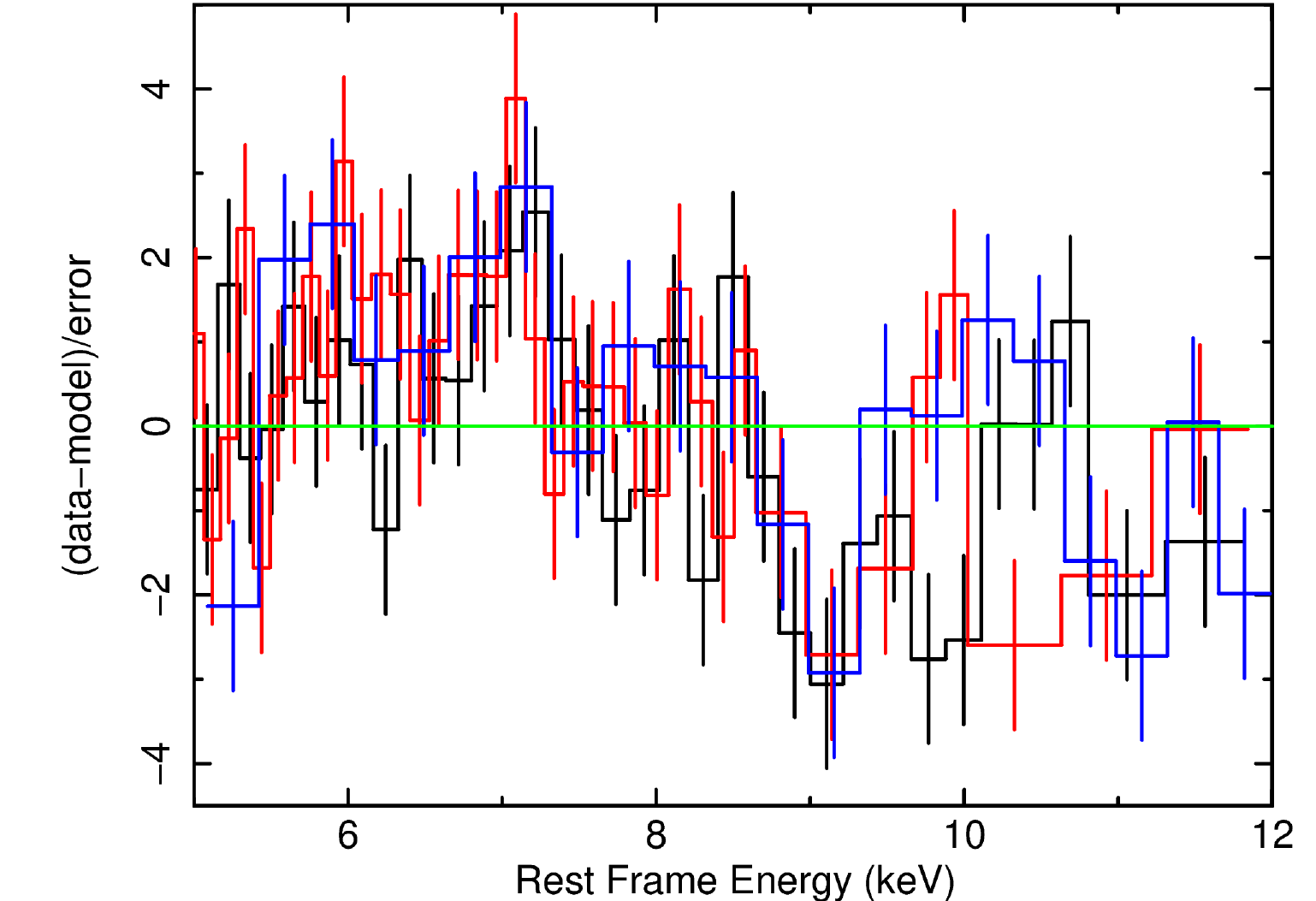}
\caption{Zoom in the soft (left) and hard (right) band of the residuals from panel b of \autoref{fig:broadb}. The spectra in the right panel have been rebinned for clarity. Colors as in \autoref{fig:broadb}.}
\label{fig:zoomSH}
\end{figure*}

We also tested the non-relativistic reflection model \xillver, part of the \relxill package \citep{Garcia14,Dauser14}, which includes a cutoff power law and its reprocessed emission (continuum and self-consistent emission and absorption features) by a distant ionized medium. We fixed the Fe abundance to Solar and the inclination of the system to $30^\circ$, the default value. Given the lack of relativistic effects from the inner disk in this model, however, there is little dependency on the emitted spectrum with inclination. 
The best fit is found for a significantly ionized reflector ($\logxi=3.4\pm0.6$), leading to slightly larger values of the photon index ($\Gamma=2.09\pm0.03$), the high energy cut-off ($\ecut=36_{-6}^{+7}$ keV) and non-negligible reflection ($R=0.34_{-0.25}^{+0.14}$). 
The emission lines from ionized Fe only partly model the emission observed at $\sim7$ keV.
The best fit shows only a marginal improvement of \dc$=5.5$ for one more free parameter (the ionization parameter of the reflector).
For consistency and a more straightforward comparison with results from the literature, in the following, we use the \pexrav model (\autoref{tab:cont}) as the baseline for the continuum fit. We stress, however, that the results on the emission and absorption features discussed below are not affected by which of the two reflection models is used.

\subsection{Spectral scan} 
\label{subsec:spec_scan}

Although both \pexrav and \xillver can broadly reproduce the observed continuum shape, they do a poor job in terms of the spectral details. This is confirmed by the goodness of fit computed by \xspec: a simple reflection model can be rejected at $95-97\%$ confidence.   
As shown in the zoom-in plots in the soft and hard band in \autoref{fig:zoomSH} (left and right, respectively), substantial positive and negative residuals with respect to the \pexrav model described above are visible. The negative residuals at $9-11$ keV rest frame appear at similar energies as the broad absorption feature reported in \citetalias{Tombesi15}. The positive and negative residuals around $1.1-1.3$ keV instead indicate emission and absorption features never reported for this source. 

To estimate the energy, width, and significance of any absorption or emission feature relative to the underlying continuum model, we ran a blind search simultaneously in all \xmm and \nustar spectra. 
The search was performed with our baseline continuum model plus a narrow Gaussian line (width fixed to $\sigma_{\rm width}=10\ev$, i.e., unresolved). We let the power-law photon index and normalization 
free to vary during the search. The reflection parameter and high-energy cut-off were frozen to the best-fit value to speed up the process. These have, however, a minor impact on the significance of narrow features such as the one investigated here, mainly thanks to the fact that the observed reflection is negligible (see the opposite case of APM~08279+5255 in \citealt{Bertola22}).

We run the scan around the energies where residuals are more prominent, i.e.,  in the soft ($0.8-1.7$ keV) and the hard ($5-12$ keV) rest frame energy band in  $50$ and $150$ energy steps, respectively. The normalization of the Gaussian component is free to vary between negative and positive values ($\pm1\times10^{-5}$) in $150$ steps. 
We then generate contour plots of the deviation in \dc from the best-fitting continuum model as a function of line energy and intensity (\autoref{fig:spec_scan}). We refer to \citet{Miniutti06}, \citet{Tombesi10}, \citet{Gofford13} and \citealt{Matzeu23} for a similar approach.

The spectral scan confirms the presence of a highly significant positive residual at $\sim1.1$ keV rest-frame (\autoref{fig:spec_scan}, left), with a blueshift with respect to the most prominent lines expected in this region of the spectrum (namely ionized Ne and Fe lines) of the order of $\vout=3-4\times10^4$ $\kms$. We also observe a strong absorption feature at higher energies, drawing a clear P-Cygni-like profile. 
If associated with the most prominent NeIX-X (at 0.91-0.92 and 1.02 keV rest-frame) and FeXVIII-XX (at 0.87 and 0.97 rest-frame) lines, the outflow velocity implied by the absorption feature would be in the range $\vout=0.24-0.33c$. The association with Fe L edges at 0.71-0.85 keV or with the blend of photoexcitation lines forming an unresolved transition array (UTA) at similar energies 
\citep{Behar01,Sako01} would imply an even larger outflow velocity.
These residuals are reminiscent of what observed in other similar targets like PDS456 \citep{Reeves16,Reeves20rgs} and PG~1211+143 \citep{Reeves18PG1211}, where, thanks to high-resolution spectra from \xmm RGS, similar emission and absorption features are resolved into multiple Fe and Ne highly ionized lines blueshifted by outflow velocities of $\sim0.26c$ and $\sim0.06c$ respectively, or IRAS~13224-3809 where Ne/Fe blended lines with outflow velocity of $\sim0.25c$ are detected through principal component analysis \citep{Parker17PCAIRAS13224}.

In the hard band (\autoref{fig:spec_scan}, right),
two distinct, highly significant absorption features are observed around 9 and 11 keV. 
Both features can be attributed to a single outflow component with $\vout\sim0.27c$ if the 9.1 keV feature is associated with a blend of Fe XXV \hea and FeXXVI \lya (at 6.70 and 6.97 keV rest-frame) and the 11 keV feature with a blend of Fe XXV \heb and XXVI \lyb (at 7.89 and 8.27 keV rest-frame). 

We note the lack of a prominent \feka line at 6.4 keV, in agreement with the low reflection measured in Sec.~\ref{subsec:continuum}, while we observe highly significant positive narrow residuals at $\sim6$ and $\sim7$ keV. If the latter is attributed to Fe XXV/XXVI \ka lines, the derived blueshift is in the range $\vout=5-17\times10^3$ $\kms$. The presence of FeXXV/XXVI emission lines would be in agreement with the high ionization found using the \xillver model for the reflector (see Sec.~\ref{subsec:continuum}).

The $\sim6$ keV feature instead would require a substantial redshift, i.e., inflow velocity of up to $3-4\times10^4$ $\kms$ if associated with highly ionized Fe. Alternatively, the whole feature can be interpreted as a broad emission feature modified by self-absorption from low-velocity ionized components, as often observed in UV and soft X-ray lines \citep{costantini07,Pounds11}.

We also note negative residuals from pn and MOS1+2 around $\sim10$ keV rest-frame. These are, however, at different energies in the two sets of EPIC cameras and not seen at all by \nustar. The difference in energy between the pn and MOS1+2 features is of the order of $\sim200$ eV, which is much larger than the energy cross-calibration uncertainties expected between the two types of \xmm cameras (a few tens of eV at these energies, R. Saxton \& M. Stuhlinger private communication). Therefore, we conclude that these are most likely spurious features. Indeed, by running the spectral scan simultaneously on all the spectra, as explained above, we do not find any significant feature at $\sim10$ keV rest-frame.

\begingroup
\renewcommand{\arraystretch}{1.10} 
    
 \begin{table}[t]
 \centering
 \caption{Best-fit parameters and significance of the Gaussian emission and absorption lines.}
   \begin{threeparttable}
    \begin{tabular}{llcc}  
        \hline
        \noalign{\smallskip}
        Component & Parameter      &  & Units \\
        \noalign{\smallskip}
        \hline
        \noalign{\smallskip} 
  {\texttt{Em$_1$}\xspace} ~ ~ ~  &   Energy  ~  ~ ~   ~    & ~ ~ ~ $1.12\pm0.06$ ~ ~ ~  &   keV \\
             &   \sw          &  $<282$            &         eV   \\
             &   EW           &  $119\pm65$      &        eV     \\
             &  \dc         &   $37.96$          &        -      \\
             &  \pft          &   $>99.99\%$     &        -      \\
             &  \pmc          &    $>99.9\%$       &      -      \\
        \noalign{\smallskip}
        \hline
        \noalign{\smallskip} 
   {\texttt{Abs$_1$}\xspace}    &   Energy     &  $1.30\pm0.02$    &       keV \\
             &   \sw         &  $<32$          &         eV   \\
             &   EW          &  $-31\pm11$      &        eV       \\
             &  \dc        &   $18.56$      &        -      \\
             &   \pft        &    $99.97\%$     &        -        \\
             &  \pmc         &    $99.8\%$      &      -         \\
        \noalign{\smallskip}
        \hline
        \noalign{\smallskip} 
   {\texttt{Em$_2$}\xspace}    &   Energy       &  $6.00\pm0.04$   &       keV \\
             &   \sw          &  $<102$          &         eV   \\
             &   EW           &  $75_{-25}^{+31}$ &       eV       \\
             &  \dc         &   $22.28$        &        -      \\
             &  \pft          &   $>99.99\%$     &        -        \\
             &  \pmc          &     $>99.9\%$    &      -         \\
        \noalign{\smallskip}
        \hline
        \noalign{\smallskip} 
   {\texttt{Em$_3$}\xspace}    &   Energy       &  $7.09\pm0.04$    &       keV \\
             &   \sw          &  $<132$            &      eV   \\
             &   EW           &  $116_{-34}^{+25}$ &       eV       \\
             &   \dc         &   $31.74$         &        -      \\
             &  \pft          &    $>99.99\%$      &       -        \\
             &  \pmc          &      $>99.9\%$    &      -         \\
        \noalign{\smallskip}
        \hline
        \noalign{\smallskip} 
   {\texttt{Abs$_2$}\xspace}   &   Energy   &  $9.13_{-0.10}^{+0.13}$ &    keV     \\
             &   \sw        &  $230_{-130}^{+180}$  &   eV   \\
             &   EW         &  $-240_{-80}^{+50}$   &   eV       \\
             &  \dc       &   $32.77$             &        -      \\
             &  \pft        &    $>99.99\%$       &        -        \\
             &  \pmc         &    $>99.9\%$       &      -         \\
        \noalign{\smallskip}
        \hline
        \noalign{\smallskip} 
{\texttt{Abs$_3$}\xspace}  &   Energy          &  $11.00_{-0.07}^{+0.06}$    &   keV \\
         &   \sw             &  $<240$   &         eV   \\
         &   EW              &  $-230_{-90}^{+70}$  &     eV       \\
         &  \dc            &  $16.59$           &    -      \\
         &  \pft             &    $99.94\%$      &        -        \\
         &  \pmc             &     $99.8\%$       &      -         \\
        \noalign{\smallskip} 
        \hline
        \noalign{\smallskip} 
            &  $\cstat/\nu$            & $392.47/363$        &  -    \\ 
\noalign{\smallskip}
       \hline
    \end{tabular}
    \begin{tablenotes}[para,flushleft]
    We report Energy, line width (\sw), and equivalent width (EW) for each emission and absorption line; $\cstat/\nu$ is the fit statistic over the number of degrees of freedom; \dc is the improvement in the fit statistic for each added component; \pft is the significance of each added component from F-test and \pmc is the significance based on Monte Carlo simulations (see \autoref{sec:line_sign} and \autoref{apsec:montecarlo} for details).
    \end{tablenotes}
    \end{threeparttable}
 \label{tab:gauss}
 \end{table}
\endgroup

As shown in \autoref{fig:broadb} (b), there are less prominent negative residuals between 2 and 5 keV. A spectral scan in this band reveals absorption features at $\sim1.9$, 2.7, and 3.4 keV rest frame at a significance of $\sim90-99\%$ confidence level. These are tentatively associated with Mg XII, Si XIV, and S XVI \lya lines (at 1.47, 2.00, and 2.62 keV rest-frame respectively), assuming a blue shift of $\sim0.27c$ (see \autoref{apsec:scan}) as derived above for the $9-11$ keV features. Indeed, modeling the hard band features with a single \xstar table as discussed in Sec.~\ref{sec:UFOmodel} will also fit and remove these residuals. 

Finally, there are noticeable negative residuals at $\sim14.6$ keV. Since only the \nustar spectra are available at these energies, the significance of the feature is quite low ($\sim95\%$ confidence level from the spectral scan). If associated again with a blend of Fe XXV/XXVI \ka, the observed energy would imply an unprecedented outflow velocity of $\vout\sim0.63c$, similar to the extreme wind velocity found in some of the observations of APM 08279+5255 \citep{Chartas09} and more recently in NGC 2992 \citep{Luminari23ApJ...950..160L}. The presence in IRASF11119 of a beamed radio jet with outflow velocity estimated to be $>0.57c$ \citep{Yang20} opens up the interesting possibility that we may observe signatures due to outflowing jet material, as tentatively reported in a handful of sources but never confirmed
\citep{Yaqoob99,Wang05,Zheng08}, and suggested to explain the 
fastest UFOs seen in APM 08279+5255 \citep{Zheng08}. 
A thorough reanalysis of the \suzaku and \nustar spectra from \citet{Tombesi15} and \citet{Tombesi17} to explore the persistence and variability of all the features discussed above will be presented in Paper II.

\section{Gaussian line properties and significance}
\label{sec:line_sign}

To constrain the properties - energy, width, intensity - and the significance of the putative emission and absorption features identified in Sec.~\ref{subsec:spec_scan}, we model each line with a Gaussian component, either in emission or absorption. The best-fit parameters of each Gaussian, labeled Em$_i$ or Abs$_i$ for emission and absorption, respectively, and ordered by increasing energy, are reported in \autoref{tab:gauss}. 
The global fit statistics of the model comprising three emission and three absorption Gaussian lines, added to the broadband continuum, is $\cstat=392.47$ for 363 d.o.f. with a significant improvement on the fit quality (see \autoref{fig:broadb} (c)). Adding Gaussian emission and absorption lines to model the main residuals identified in Sec.~\ref{subsec:spec_scan} does not significantly impact the continuum best-fit values, which are therefore not reported in \autoref{tab:gauss}.

The location of the Gaussian lines is well constrained, with errors on the line energy of a few $\%$ in all cases.
The width of the lines $\sigma_{width}$ is consistent with a narrow, unresolved line in all cases except for the absorption line at 9 keV ($\sigma_{width}=232_{-132}^{+184}$ eV). When the line is consistent with being unresolved, we freeze the $\sigma_{width}$ to its best-fit value.
The equivalent width (EW) of the emission lines is in the range $EW\sim75-120$ eV, while the absorption lines at 9-11 keV have $EW\sim-230$ eV. The absorption line in the soft is, instead, shallower with $EW\sim-30$ eV.

As a first assessment of the line significance, we compute the \dc between the best fit with and without each Gaussian line and the F-test null hypothesis probability considering two more free parameters (energy and normalization) for all cases except the absorption line at 9 keV, where also the width of the line is a free parameter. The improvement in the fit statistic \dc is reported in \autoref{tab:gauss} for each emission or absorption line. \pft, defined as $1-P_{null}$ where $P_{null}$ is the null hypothesis probability computed from the F-test, is also reported.
All the lines are highly significant, with \pft$>99.9\%$, the weaker one being the feature at 11 keV, as also shown by the contour plots in 
Sec.~\ref{subsec:spec_scan}. It is, however,  well known that deriving the significance of additional components using the F-test is not appropriate when looking at emission/absorption lines \citep{Protassov02, Markowitz06}, as the null model is at the boundary of the parameter space (i.e., line normalization equal to 0). Furthermore, the results from the F-test tend to overestimate the significance of these components as they do not consider the full range of energies where a line might be observed by chance, the so-called {\it look-elsewhere effect}.

For this reason, we carried out a more rigorous test to assess the significance of the line based on Monte Carlo ($\mathcal{MC}$) simulations (see, e.g., \citealt{Porquet04monte, Miniutti06,Tombesi10,Gofford13,Matzeu23} for a similar approach). 
The details of the simulations can be found in \autoref{apsec:montecarlo}. 
The goal of such simulations is to derive the significance of each added component \pmc, defined as \pmc$=1-\left(\frac{N}{S}\right)$ where N is the number of simulated spectra showing a random fluctuation larger than the one observed in the data and S the total number of simulated spectra.
The values of \pmc derived with this approach are reported in \autoref{tab:gauss}. None of the 1000 realizations in the soft and hard bands show a \dc greater than the one observed for Em$_{1,2,3}$ and Abs$_2$
leading to \pmc$>99.9\%$ for all these features, while \pmc$=99.8\%$ for the remaining two absorption features.

\section{Ultra Fast Outflow modeling}

\begin{figure*}
\centering
\includegraphics[width=0.48\textwidth]{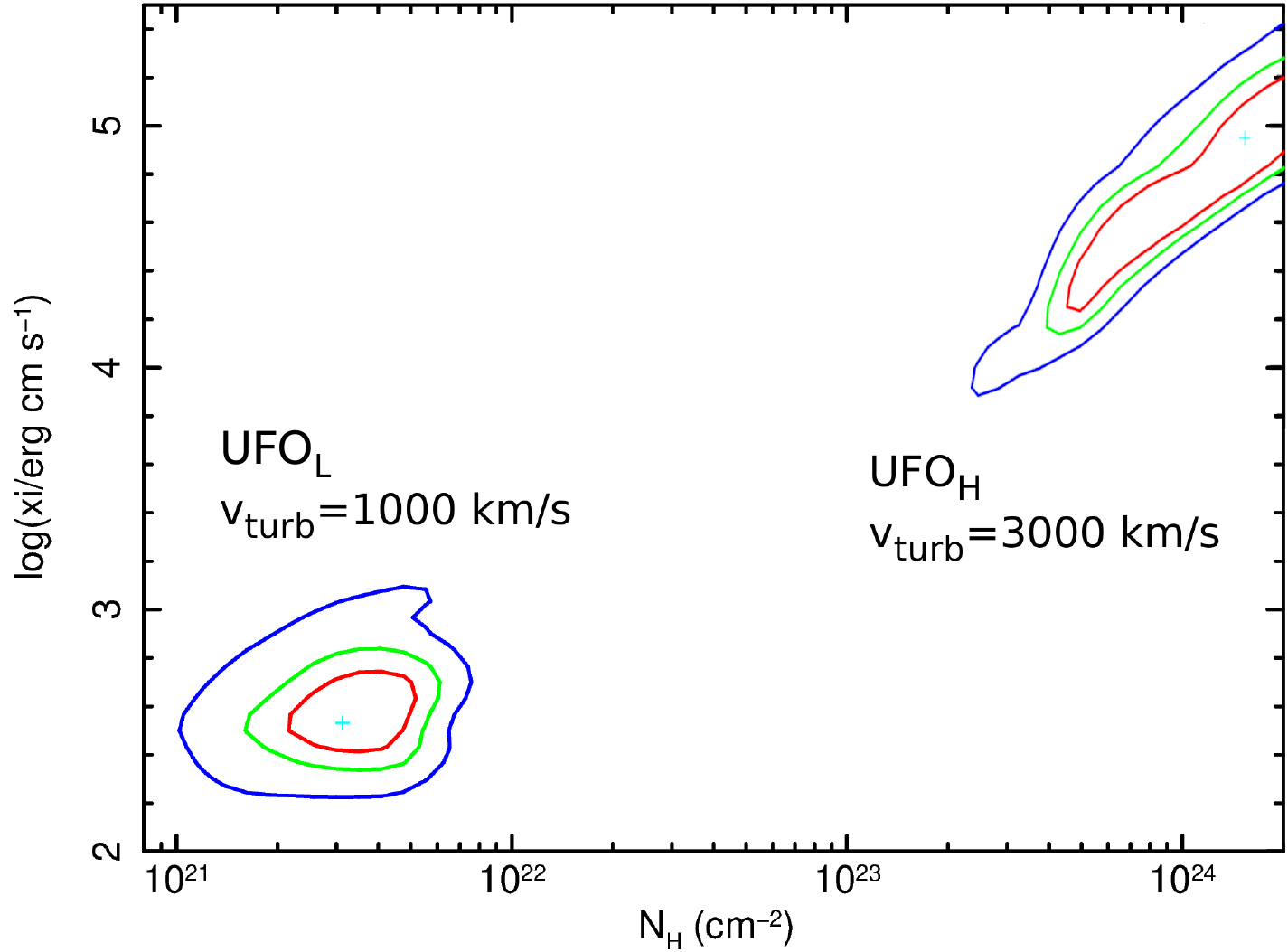}\hspace{0.5cm}
\includegraphics[width=0.48\textwidth]{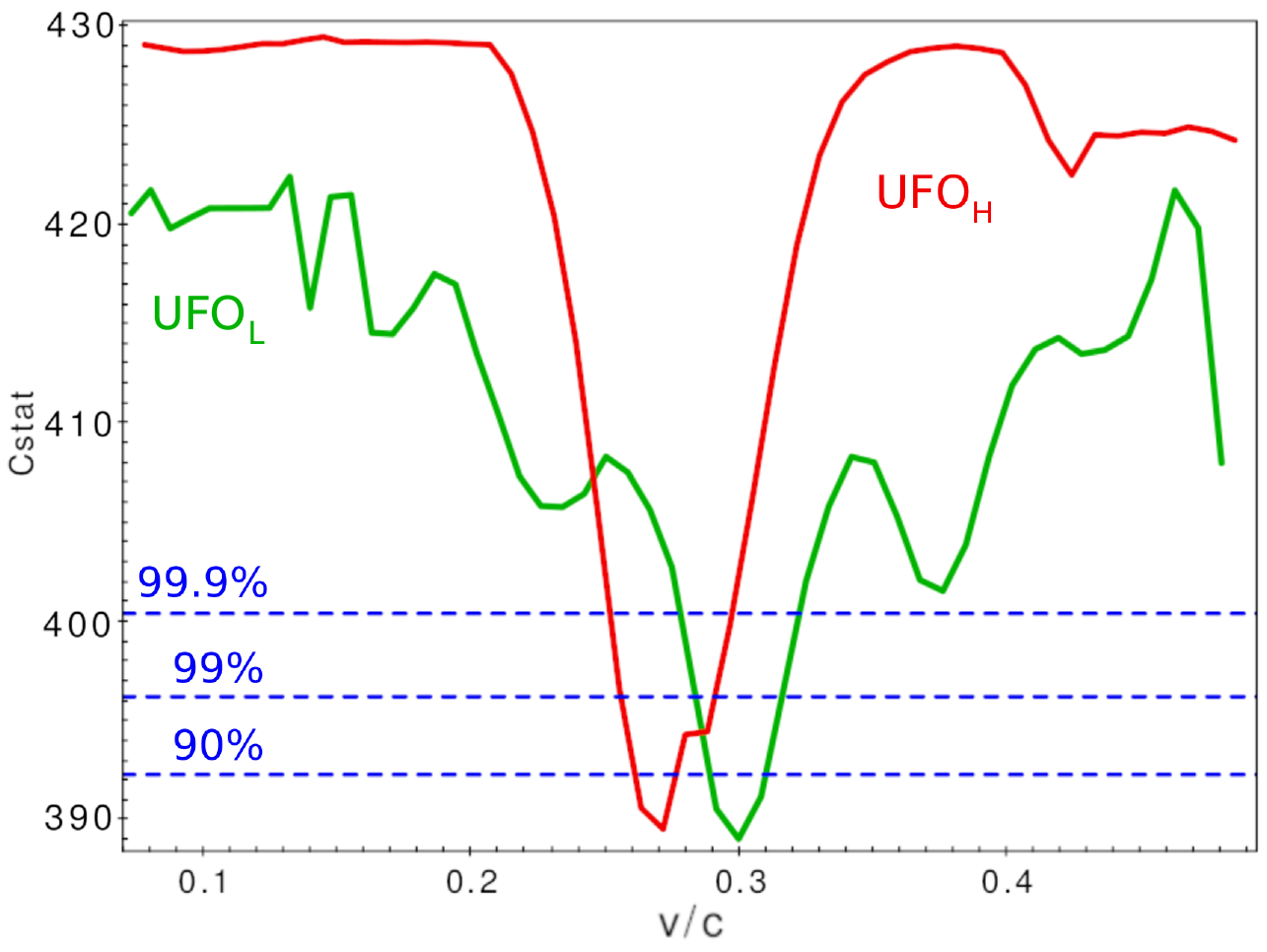}
\caption{Left: Confidence contour levels of the column density $\nh$ and the ionization parameter $\logxi$, for the best-fit with two \xstar tables, one with $\vturb=1000\kms$ for the low-ionization table {UFO$_{\rm L}$} and one with $\vturb=3000\kms$ for the high-ionization {UFO$_{\rm H}$} one. Right: Contour plot of the outflow velocity for the \ufoh in red and for \ufol in green. The blue lines mark one parameter's 90, 99, and 99.9\% confidence levels.
}
\label{fig:cont_nhxi}
\end{figure*}

\label{sec:UFOmodel}
\begingroup
\renewcommand{\arraystretch}{1.4} 
 \begin{table}[t]
 \centering
 \caption{Continuum and \xstar absorption tables best-fit parameters. One \xstar table with $\vturb=300\kms$ is used for the WA, one with $\vturb=1000\kms$ for the soft UFO, and one with $\vturb=3000\kms$ for the hard UFO.}
   \begin{threeparttable}
    \begin{tabular}{llcc}  
        \hline
        \noalign{\smallskip}
        Component & Parameter      &  & Units \\
        \noalign{\smallskip}
        \hline
        \noalign{\smallskip} 
    \pexrav ~ &   $\Gamma$   ~ ~     & ~ $2.05\pm0.06$  ~   & ~  - ~ \\ 
            &    R               & $<0.35$            &  -    \\
            &    $\ecut$          & $25.4_{-5.0}^{+9.5}$  &  \kev \\
            &    $f_{scat}$        &  $1.2\pm0.4\%$        &  -    \\   
 {\texttt{WA}\xspace}  &   log($\nh$)  &  $22.43_{-0.04}^{+0.06}$     &  $\cmsq$    \\ 
            &   log($\xi$)          &  $0.05_{-0.17}^{+0.25}$    &  erg cm/s    \\
            &   $\vturb$          & $300^*$           &  km/s  \\
            &  $\vout$            & $<2500$           &  km/s  \\
 {\texttt{UFO$_{\rm L}$}\xspace}  &   log($\nh$)       &  $21.54_{-0.23}^{+0.26}$     & $\cmsq$   \\ 
            &  log($\xi$)          &  $2.57_{-0.17}^{+0.26}$      &  erg cm/s  \\
            &   $\vturb$          & $1000^*$           &  km/s  \\
            &  $\vout$            & $0.30\pm0.01$      &  v/c  \\
{\texttt{UFO$_{\rm H}$}\xspace}  &   log($\nh$)       &  $>23.99$     &  $\cmsq$   \\ 
            &   log($\xi$)           &  $4.95_{-0.45}^{+0.43}$    &  erg cm/s   \\
            &   $\vturb$           & $3000^*$           &  km/s  \\
            &   $\vout$            & $0.27\pm0.01$      &  v/c  \\
        \noalign{\smallskip} 
    \hline
        \noalign{\smallskip} 
            &  $\cstat/\nu$            & $389.3/369$  &  -  \\
            &   $L_{\rm 2-10\,keV}$   & $1.3\times10^{44}$  &  $\ergs$  \\  
        \noalign{\smallskip} 
    \hline
    \end{tabular}
    \begin{tablenotes}[para,flushleft]
    \pexrav parameters are defined as in \autoref{tab:cont}; log($\nh$), log($\xi$), $\vturb$ and $\vout$ for WA, {\texttt{UFO$_{\rm L}$}\xspace} and  {\texttt{UFO$_{\rm H}$}\xspace} components are defined as in \autoref{tab:cont}; $\cstat/\nu$ is the fit statistic over the number of degrees of freedom;
    $L_{\rm 2-10\,keV}$ is the absorption-corrected luminosity in the 2-10 keV rest frame band; $*$~fixed parameters.
    \end{tablenotes}
    \end{threeparttable}
 \label{tab:xstar}
 \end{table}
\endgroup

In the following, the same continuum model and emission lines described above are adopted, while the absorption lines are replaced by absorption tables from two different outflow templates. The continuum and emission line best-fit parameters are re-fitted to accommodate changes in the global fit. 

\subsection{\xstar tables}
\label{subsec:xstar}

As a first step, we adopt absorption tables created with the \xstar photo-ionization code \citep{BautistaKallman01,Kallman04}. This allows us to constrain the physical properties of the absorbing medium and, in particular, the ionization parameter $(\xi)$, the column density $(\nh)$, and the outflow velocity ($v_{\rm out}$). 

We start from the broadband continuum model described in Sec.~\ref{subsec:continuum}. The \xstar table modeling the WA and the three emission lines described in Sec.~\ref{sec:line_sign} are kept in the model, while the three absorption lines are removed and replaced by two \xstar tables, as both the features at $\sim$9 and $\sim$11 keV can be modeled with the same highly ionized dense gas layer. We used the tables created with the \xstar suite v2.54a for the SUBWAYS sample \citep{Matzeu23}, which cover a wide range of column densities ($\lognh=22-24.3$) and ionization parameters ($\logxi=2-8$). These were generated for power-law SED of $\Gamma=2$ input spectrum and an ionizing luminosity of $\lion=1\times10^{45}$ $\ergs$, well matched with the values observed in IRASF11119.

One of the input parameters needed to create the tables is the turbulent velocity $\vturb$.
To determine which input $\vturb$ is better suited to fit the observed features, we tested different 
tables with input $\vturb$ in the range $10^3-10^4 \kms$. 
We found that $\vturb=3000 \kms$ and $\vturb=1000 \kms$ reproduce best the observed hard band and soft band features, respectively. Therefore, in the following, we use the pre-compiled tables with such velocities. 
Note that these values are lower than what can be derived at face value from the $\sigma$ best-fit values of the Gaussian lines reported in \autoref{tab:gauss} ($\sim8000 \kms$).
This is because, in that approach, a single Gaussian profile is used to fit what is, in reality, the blend of multiple features: Fe XXV/XXVI \hea and \lya in the case of the 9 keV feature, and several Fe and Ne lines in the case of the 1.3 keV feature.

One absorption table has a high best-fit column density and ionization parameter, as these are needed to reproduce the two main features in the hard band through a blend of Fe XXV/XXVI \hea-\lya and \heb-\lyb transition pairs. 
The outflow velocity is $\vout=0.27\pm0.01$c, falling in the Ultra Fast Outflow regime. 
We labeled this component as \ufoh as it represents the high-ionization UFO component.

The soft feature instead is reproduced by a relatively low column and low ionization gas layer, with best-fit values typical of warm absorbers \citep{Piconcelli05,McKernan07}. The outflow velocity, $\vout=0.30\pm0.01$c is, however, much higher than the typical WA and well within the UFO regime.
We labeled this component as \ufol as it represents the low-ionization counterpart of the UFO.
The two velocities are statistically self-consistent, as shown in \autoref{fig:cont_nhxi}, right, suggesting that the two components are part of the same outflow. 
The best-fit parameters of the two \xstar tables are shown in \autoref{tab:xstar}, while \autoref{fig:cont_nhxi}, left, shows the confidence contours of log($\nh$) vs. log($\xi$) for the two tables, while the residuals with respect to this model are shown in \autoref{fig:broadb} (d).

The column density of the \ufoh component is at the limit of the range explored by the \xstar tables adopted, up to $\nh=2\times10^{24} \cmsq$. Above this value, the derived intrinsic luminosity would be unreliable
as \xstar does not take into account Compton scattering.
Log($\nh$) and log($\xi$) of the highly ionized gas are also strongly correlated, as higher ionization states produce weaker absorption troughs that, in turn, can accommodate larger columns to fit the same observed features.

\begin{figure}
\centering
\includegraphics[width=0.49\textwidth]{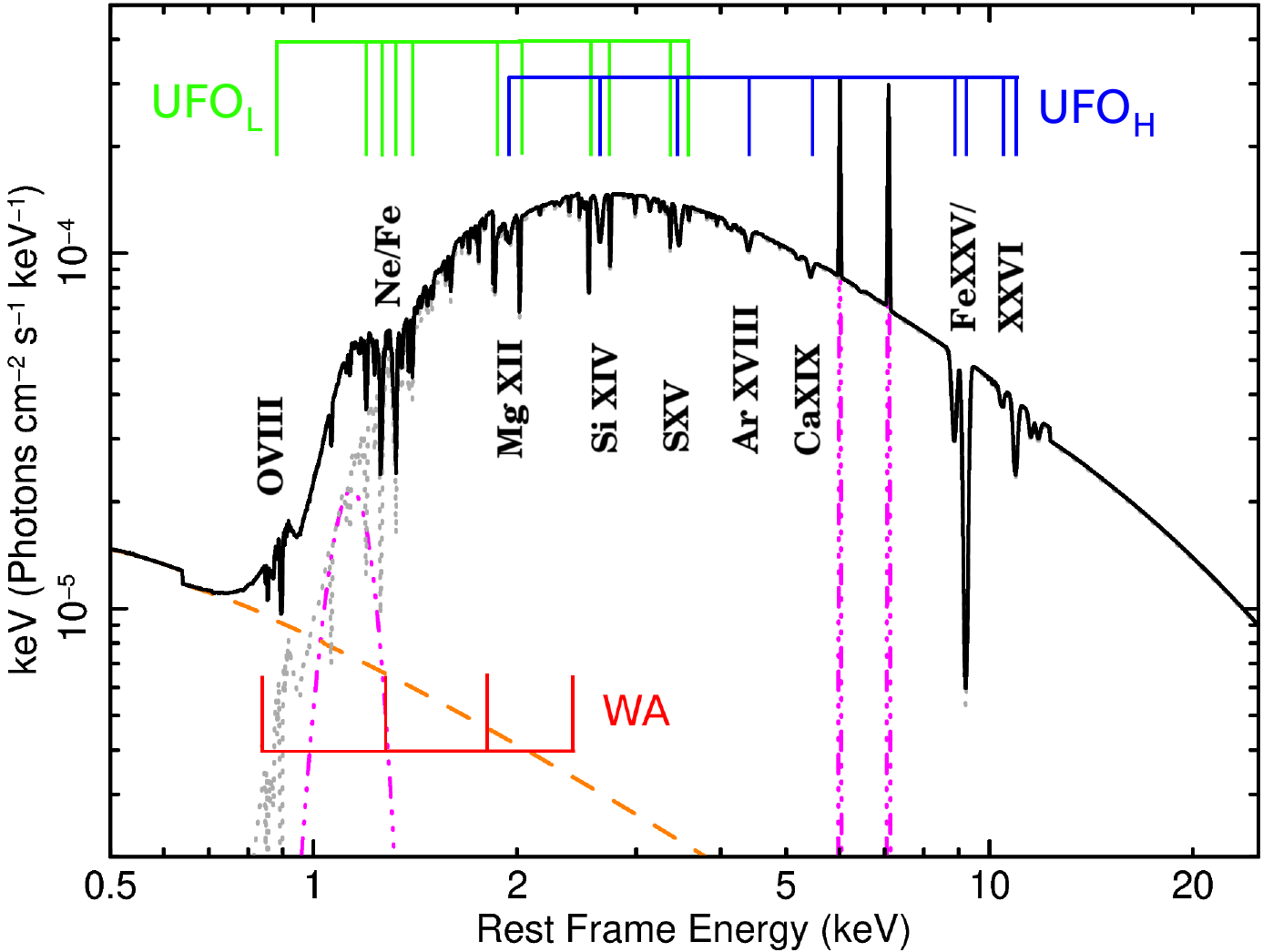}
\caption{
Unfolded best-fit model (black) in $\nu F(\nu)$ including three \xstar absorption tables. The emission lines are shown in magenta, and the scattered emission is in orange. The main absorption features due to the WA, \ufol, and \ufoh are labeled in red, green, and blue, respectively. The ions producing the most prominent features are labeled in black.}
\label{fig:bestfit}
\end{figure}

The outflow velocities of both \ufol and \ufoh are well within the UFO regime ($\vout>10^4\kms$) and on the high-velocity tail of the distributions observed in the local Universe \citep{Tombesi10,Gofford13,Igo20} and up to z=0.5  \citep{Matzeu23}: there are only 3 cases of UFO with faster velocities among the 65 individual detections in the spectra of the four combined samples, with several sources in common. In \autoref{fig:bestfit}, we show the best-fit model with the three \xstar tables described above. We labeled in red the main absorption features produced by the WA table, in green those relative to the \ufol table, and in blue those relative to the \ufoh table. The ions producing the main features are also labeled.  

We stress that only one \xstar table is needed to model both the 9 and 11 keV features, and the same table also models the less significant residuals observed at 2, 2.7, and 3.4 keV (see \autoref{apsec:scan}). As a result, the \dc obtained when adding this component is \dc$\sim42$ for three more free parameters. This is significantly larger than the \dc associated with each of the Gaussian lines reported in \autoref{tab:gauss}, boosting the overall significance with respect to the results derived in Sec.~\ref{sec:line_sign}. Adding the \xstar table for the \ufol gives a \dc$\sim15$, comparable with the value derived for a simple Gaussian line.  The global fit statistics of the model comprising three emission Gaussian lines and three \xstar absorption tables for WA, \ufol and \ufoh is $\cstat=389.3$ for 369 d.o.f.

\begin{figure*}
\includegraphics[height=9.0cm,width=9cm]{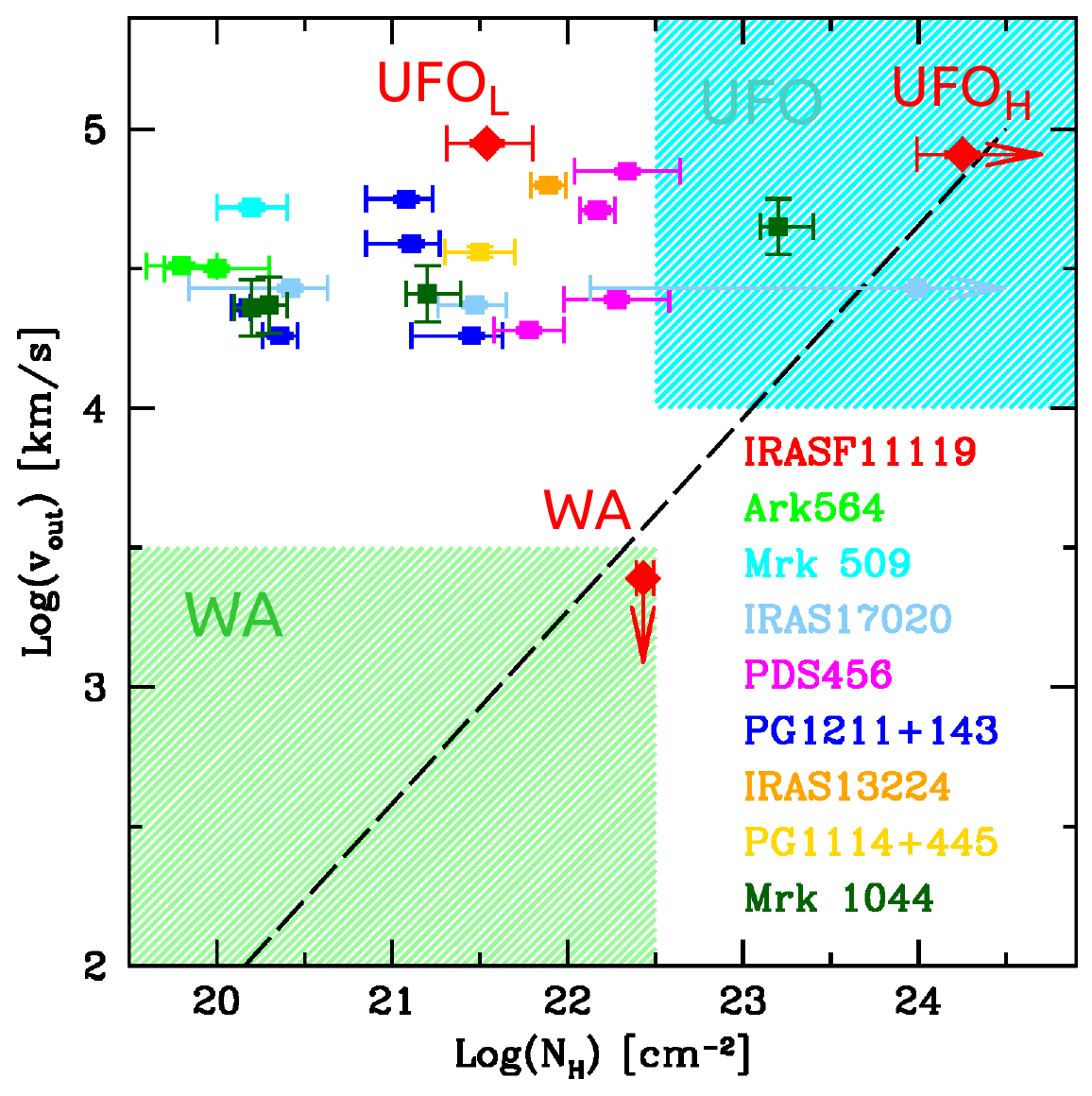}\hspace{0.5cm}
\includegraphics[height=9.0cm,width=9cm]{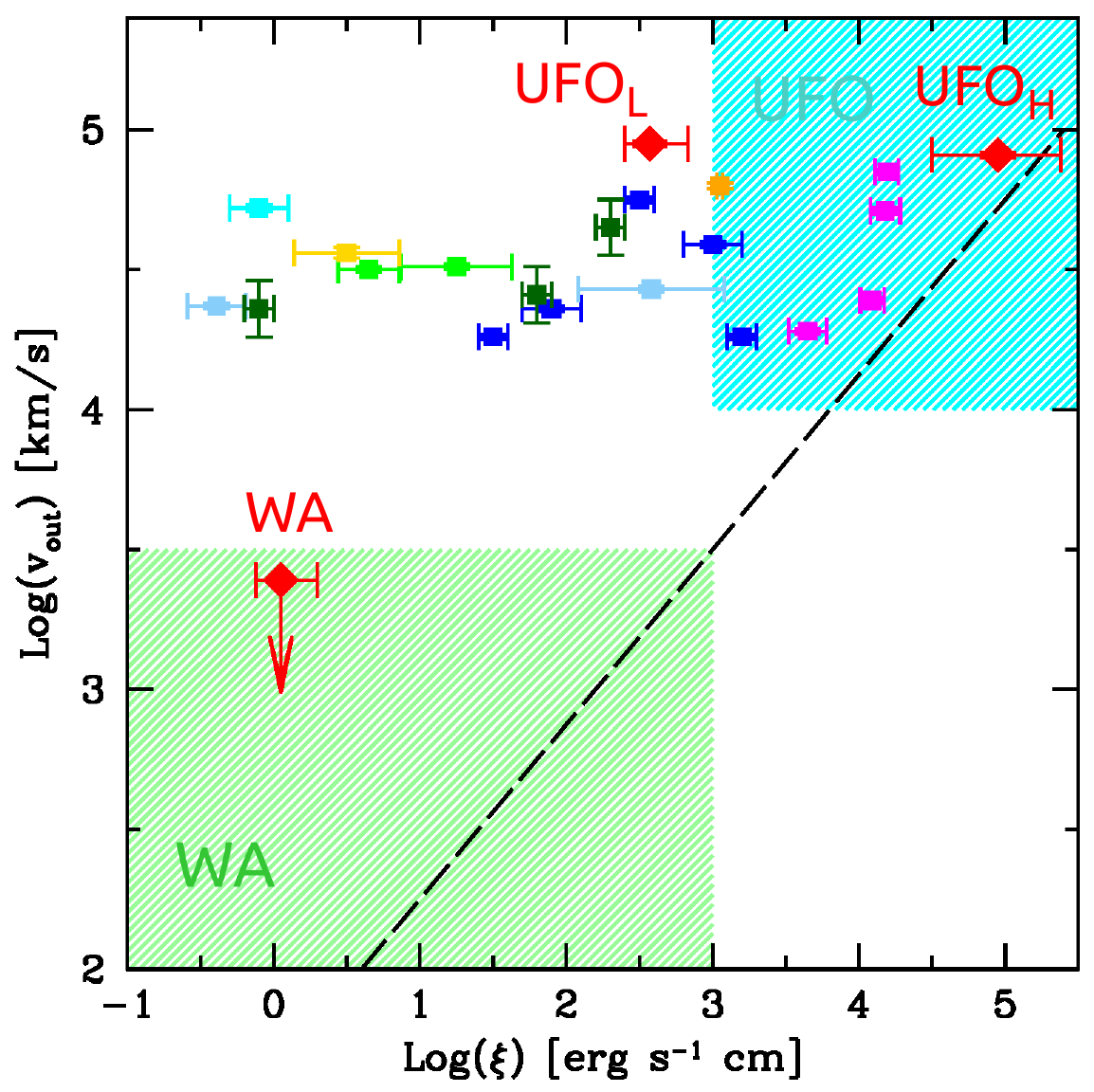}
\caption{log($\nh$) vs. $\vout$ (left) and log($\xi$) vs. $\vout$ (right) for the WA, \ufol and \ufoh components of IRASF11119 (red diamonds). The known soft-band detected fast winds from the literature are shown with colored points: light green \citet{Gupta13b}; cyan \citet{Gupta15};
sky blue \citet{Longinotti15}; magenta \citet{Reeves16}; blue \citet{Pounds16}; orange \citet{Jiang2018MNRAS.477.3711J}; yellow \citet{Serafinelli19}; dark green \citet{Krongold21}. The low-velocity WAs region (green), UFOs region (cyan), and the correlation fits linking low-velocity WAs and UFOs (dashed black lines) are derived from \citet{Tombesi13}. 
}
\label{fig:nhvout}
\end{figure*}

In order to have a rough estimate of the covering fraction of the wind, we followed the  approach described in, e.g., \citealp{Nardini15} and \citealp{Reeves16}, fitting the  
excess emission in the hard band (after removing the Gaussian lines at $6$ and $7 \kev$)
with an additive table created by \xstar together with the absorption tables, that models the reflected emission from the wind.
Indeed, the reflected emission table is able to model the emission feature at $7 \kev$ with emission from the FeXXVI \lya line, whose intensity is actually driving the normalization of the whole reflected component (see \citealt{Nardini15} for a similar case on PDS456).

The normalization $k$ of the reflected component is defined by \xstar in terms of $k = f L_{38}/D^2(kpc)$ where $L_{38}$ is the ionizing luminosity in units of $10^{38} \ergs$, and D(kpc) is the distance to the quasar in kpc and $f$ is the covering fraction of the gas with respect to the total solid angle, in terms of  $f = \Omega /4\pi$.
Thus, by comparing the predicted normalization  for a fully-covering shell of gas versus the
observed normalization ($k_{xstar}$), the covering fraction of the gas can be estimated. 
For IRASF11119, with $\lion = 4\times10^{45} \ergs$ at a luminosity
distance of $D = 779$ Mpc, the observed value of the normalization ($k=5.9_{-3.1}^{+5.1}\times10^{-5}$) implies a covering fraction of $f> 0.4$.

In \autoref{fig:nhvout}, we show the distribution of log($\nh$) vs. $\vout$ (left) and log($\xi$) vs. $\vout$ (right) best-fit values of the three \xstar absorption tables adopted here and compare them with the typical UFO and WA parameter spaces (cyan and green, respectively) taken from the literature \citep[e.g.,][]{Tombesi13,Gupta13,Serafinelli19}.  While the WA and {UFO$_{\rm H}$} components fall in the respective classical regions, the outflow observed in the soft band falls in a poorly populated region of low ionization/low column clouds outflowing at velocities typical of UFOs.  

This \ufol component represents a promising new example of a poorly explored phase of the UFO phenomenon, with only about a dozen other sources known to date to show such features mainly detected in high-resolution grating spectra. These features are thought to be produced by ISM clouds at great distances from the SMBH (tens/hundreds of parsec, also referred to as `meso-scale'), entrained by the advancing outer shock of the nuclear UFO \citep[e.g.][]{Fauchere12}. 
Alternatively, they may represent the imprint of the UFO gas itself cooling down and fragmenting \citep[e.g.][]{Takeuchi13}. In the latter case, the expected distances would be much smaller, of the order of hundreds of $\rg$.
Variability of these features may help discern between the two interpretations, as in the case of PG~1211+143 \citep{Reeves18PG1211}, a thorough search for long and short-term variability of both the UFO features presented here will be addressed in Paper II. 

The growing number of highly accreting sources in which this new outflow phase has been discovered, possibly located at the meso-scale between the nuclear UFO and the larger scale WA, supports the idea that all these different wind components are, in fact, different phases of a single phenomenon expanding from the inner regions of the AGN into the host galaxy. In particular, this may be interpreted as the interface zone in which the nuclear UFO interacts with the clumpy regions at 10s-100s of pc distance, forming blobs of fast-moving but denser and colder gas. This picture agrees with predictions by models of self-regulating feedback mechanisms for the evolution of galaxies \citep{Zubova12,King15,Gaspari17_cca}.

It is interesting to note that, while in the case of `classical' UFOs at the accretion disk scale, $\vturb$ is generally interpreted as a velocity gradient within the outflow 
(with values up to 10000 km/s in the most extreme cases, \citealp[e.g.][]{Nardini15,Tombesi15}),  
for the \ufol at larger scales $\vturb$ could really be interpreted as an indication of the amount of turbulence in the flow, as 
for sub-relativistic outflows, at least several 100 km/s of turbulence can be expected at large scales \citep{Wittor20}.
Indeed, turbulence is key to having entrained outflows and thus generate different absorbers along the LOS, as found here: the stronger the turbulence, the larger the K-H instabilities, hence the entrainment \citep[e.g.][]{Gaspari20}.

\subsection{Disk Wind table}
\label{subsec:diskwind}

\begingroup
\renewcommand{\arraystretch}{1.4} 
 \begin{table}[t]
 \centering
 \caption{Best-fit parameters of the model where a \dwind grid is used to replace the \xstar table to model the {UFO$_{\rm H}$} features.}
   \begin{threeparttable}
    \begin{tabular}{llcc}  
        \hline
        \noalign{\smallskip}
        Component & Parameter      &  & Units \\
        \noalign{\smallskip}
        \hline
        \noalign{\smallskip} 
    \pexrav  &   $\Gamma$         & $2.05\pm0.05$     &  -  \\ 
            &    $R$               & $<0.35$            &  -    \\
            &    $\ecut$          & $24.3_{-4.6}^{+2.6}$  &  \kev \\
            &    $f_{scat}$        &  $0.45\pm0.4\%$        &  -    \\   
 {\texttt{WA}\xspace}  &   log($\nh$)      &  $22.54_{-0.07}^{+0.02}$     &  $\cmsq$ \\ 
            &   log($\xi$)        &  $-0.03_{-0.19}^{+0.27}$    &  erg cm/s    \\
            &   $\vturb$          & $300^*$        &  km/s  \\
            &  $\vout$            & $<2500$        &  km/s  \\
 {\texttt{UFO$_{\rm L}$}\xspace}  &   log($\nh$)  &  $21.60_{-0.25}^{+0.23}$   &  $\cmsq$  \\ 
            &   log($\xi$)          &  $2.57_{-0.24}^{+0.19}$      & erg cm/s    \\
            &   $\vturb$          & $1000^*$           &  km/s  \\
            &  $\vout$            & $0.30\pm0.01$      &  v/c  \\
\dwind      &   $\fv$             &  $1.53_{-0.11}^{+0.07}$ ($1.5^{\bf a}$)    &  -    \\ 
            &     ($\vinf$)       & ($0.38\pm0.02$)    &  v/c  \\
            &   $\mu=cos \theta$   &  $0.53_{-0.02}^{+0.03}$ ($0.525^{\bf a}$)   &  -    \\
            &   $\mw$           & $0.99_{-0.17}^{+0.26}$   &  ($\frac{\mout}{\medd}$)  \\
            &   $\lx$           & $>1.72\%$  & ($\frac{L_{\rm 2-10\kev}}{\ledd}$)  \\
        \noalign{\smallskip} 
    \hline
        \noalign{\smallskip} 
            &  $\cstat/\nu$            & $390/370$   &  -  \\
            &   $L_{\rm 2-10\,keV}\smallskip^{\bf b}$     & $5.4\times10^{44}$  &  $\ergs$  \\   
        \noalign{\smallskip} 
    \hline
    \end{tabular}
    \begin{tablenotes}[para,flushleft]
    \pexrav parameters are defined as in \autoref{tab:cont}; log($\nh$), log($\xi$), $\vturb$ and $\vout$ for WA, and {\texttt{UFO$_{\rm L}$}\xspace} components are defined as in \autoref{tab:cont};
    For the \dwind component
    $\fv$ is the velocity parameter defined as $\fv=\vinf/\vesc$;
    $\vinf$ is the corresponding terminal velocity for $\vesc=0.25$;
    $\mu$ is the cosine of the inclination angle;
    $\mw$ is the mass outflow rate in $\medd$ units;
    $\lx$ is the ionizing luminosity in $\ledd$ units. 
    $\cstat/\nu$ is the fit statistic over the number of degrees of freedom;
    $L_{\rm 2-10\,keV}$ is the absorption-corrected luminosity in the 2-10 keV rest frame band; 
    $\bf ^a$~$\fv$ and $\mu$ closest grid points used to derive more reliable constraints on $\mw$ and $\lx$. $*$~fixed parameter.
      \end{tablenotes}
    \end{threeparttable}
 \label{tab:dwind}
 \end{table}
\endgroup

To model the UFO features, we also tested the publicly available\footnote{\url{https://gabrielematzeu.com/disk-wind/}} \dwind table of simulated radiation-driven wind spectra generated using the Monte Carlo radiative transfer code developed in \citet{Sim08,Sim10,Matzeu22a}. 

The \dwind computes emission and absorption spectra arising from a smooth, steady-state, and bi-conical flow with an opening angle of 45 degrees. With such assumptions, we can provide a self-consistent treatment of both transmitted and scattered components from a 
realistic geometry that includes a radial-dependent ionization and velocity structure across the flow.
We can probe the overall mass outflow rate, velocity, and viewing angle of classical UFOs. In this work, we adopted the \dwf grid, which assumes a lunching radius of $32\rg$. The parameters of the fit are:

\begin{itemize}

    \item The velocity parameter ($\fv$), which relates the terminal velocity ($\vinf$) of the wind to the escape velocity at the origin ($\vesc=(2G\mbh/\rm R_{\rm esc})^{1/2}$) via $\fv=\vinf/\vesc$. In the case of the \dwf grid adopted here, the inner edge of the wind is launched at R$_{\rm esc}=32\rg$ and therefore $\vesc=0.25c$.

    \item The line-of-sight inclination angle ($\mu$) defined as $\mu=\cos(\theta)$ where $\theta$ is the angle between the observer's line-of-sight and the polar $z$-axis of the flow.
    
    \item The mass outflow rate ($\mw$), defined as $\mw=\mout/\medd$, i.e., normalized by the Eddington limit, making this measurement black hole mass invariant.

    \item The ionizing luminosity ($\lx$) defined as $\lx=(L_{\rm 2-10\kev}/\ledd)\times100$, which is also normalised to the Eddington limit.

    \item The photon index $\Gamma$ of the power-law continuum, extrapolated over the $0.1-511\kev$ range, used to self-consistently compute the wind spectrum. We linked this parameter to the \pexrav best-fit photon index, recomputed after including the \dwind table.  

\end{itemize} 

\noindent More details on the \dwind parameters are provided in Sec.~2 of \citet{Matzeu22a}.

\begin{figure*}
\includegraphics[height=6.2cm,width=8.5cm]{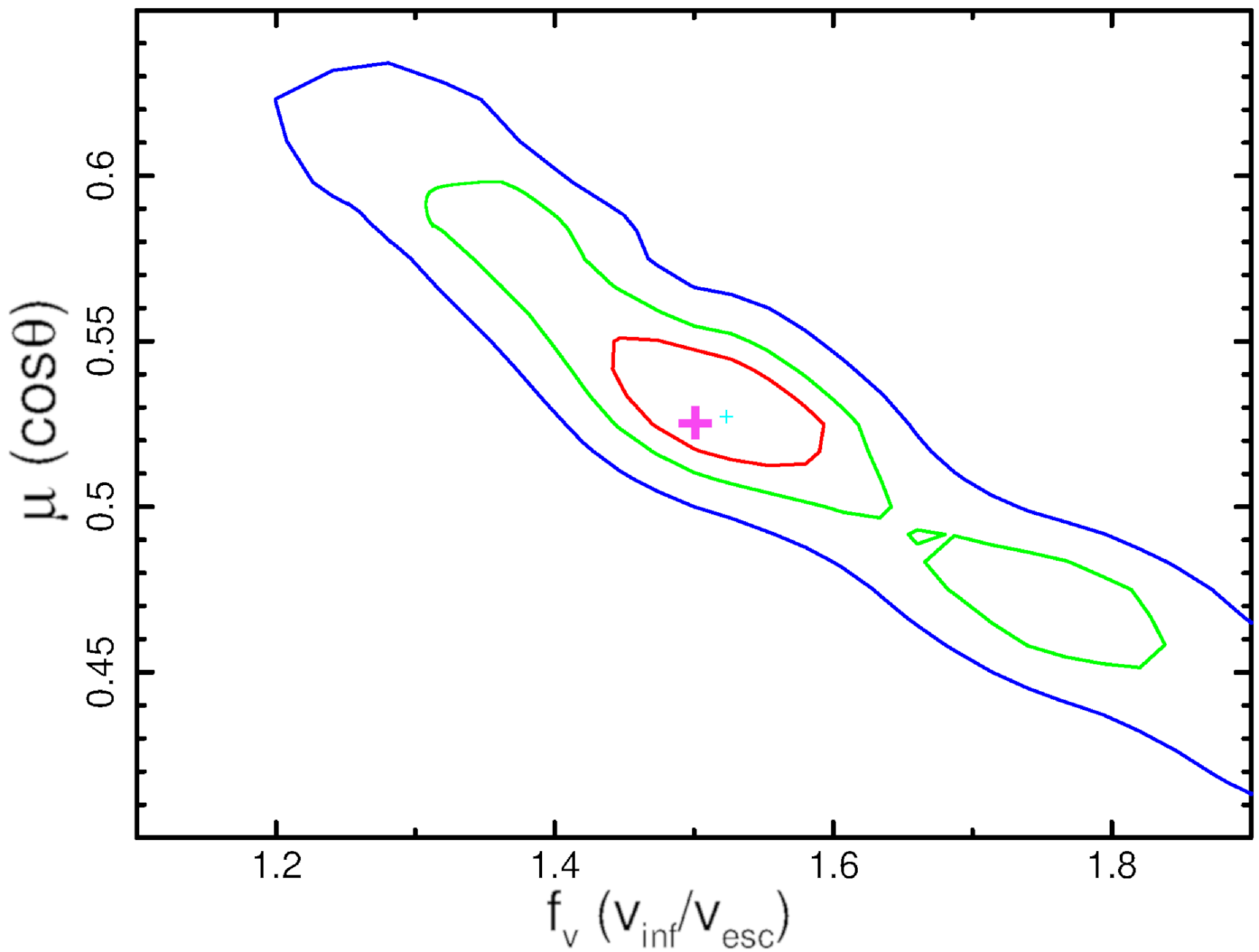}\hspace{0.5cm}
\includegraphics[height=6.2cm,width=8.5cm]{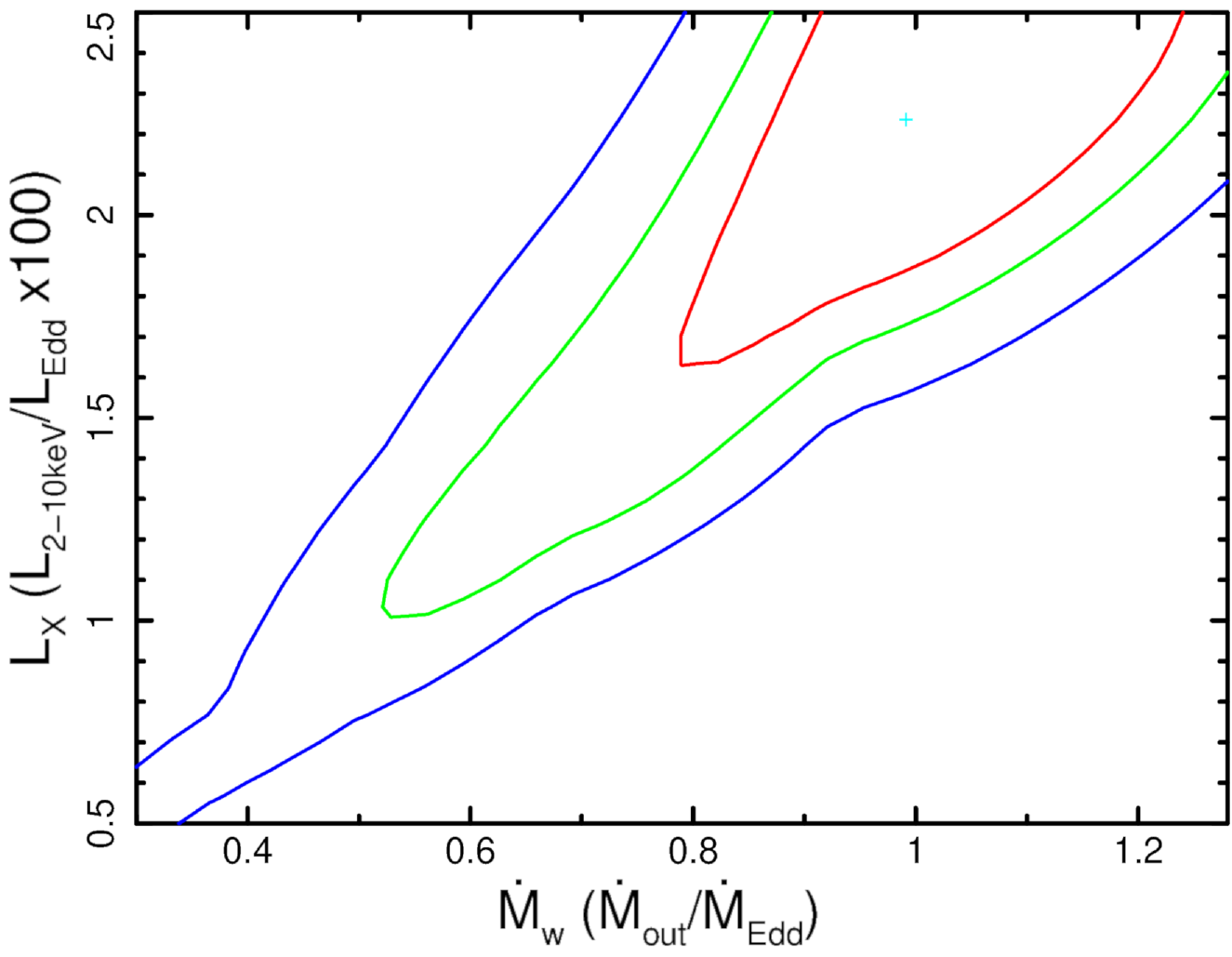}
\caption{Confidence contour levels of the \dwind parameters $\fv$ vs. $\mu$ (left) and $\mw$ vs. $\lx$ (right), see \autoref{subsec:diskwind} for their definitions. The $\mw$ vs. $\lx$ contours are obtained by fixing $\fv$ and $\mu$ to their closest grid points (marked with the magenta cross) to avoid interpolation issues.}
\label{fig:cont_dw}
\end{figure*}

Together, $\fv$ and $\mu$ regulate the observed location and width of the absorption features produced by the outflowing gas: $\fv$ by setting the terminal velocity of the flow, while the launch velocity is fixed, and $\mu$ by setting which part of the flow is intercepted by the line of sight. $\mw$ regulates the amount of material in the flow and, therefore, the depth of the absorption features. Finally, $\lx$ regulates the degree of ionization at the location where the l.o.s. intercepts the flow.

The \dwind model is designed to reproduce the typical UFO absorption features observed in the hard band (i.e., high column and high ionization state). Even if the parameters $\mw$ and $\lx$ cover a wide range of values ($\mw=[0.0196-1.287]\medd$ and $\lx=[0.025-2.516]\%\ledd$) 
these intervals do not extend enough to the lower end to cover the parameter space needed to reproduce the \ufol feature observed in IRASF11119. 
Therefore, we adopt the \dwind table to model only the \ufoh features, replacing the relative \xstar table, while we keep the rest of the model components described in Sec.~\ref{subsec:xstar}.

\begin{figure}
\includegraphics[height=6.2cm,width=8.5cm]{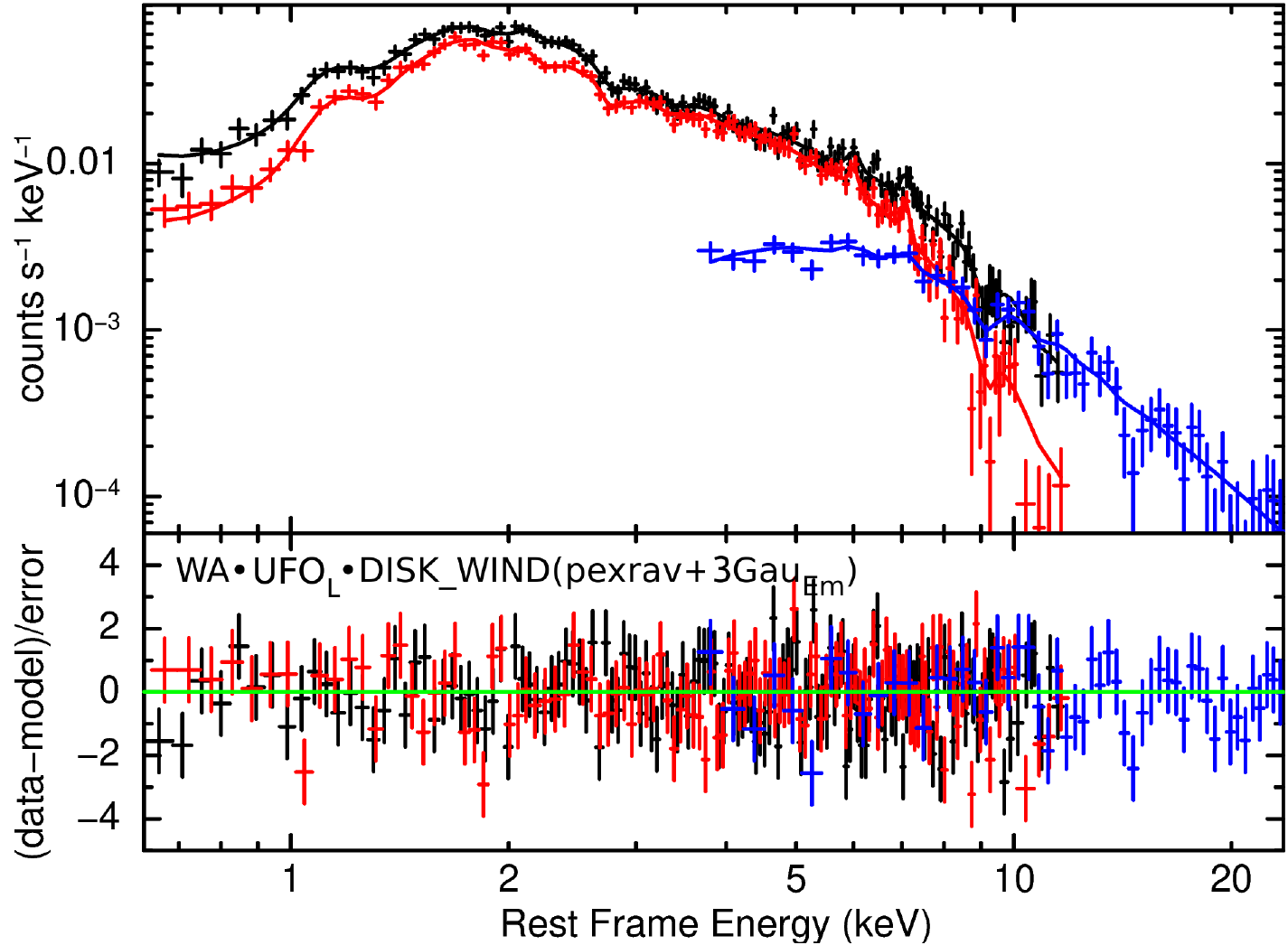}
\caption{\xmm and \nustar spectra with best-fit continuum model and residuals, for the model
described in Sec.~\ref{subsec:diskwind}, where the \dwind table replaces the \ufoh \xstar table.
Colors as in \autoref{fig:broadb}.}
\label{fig:spec_dw}
\end{figure}

The best-fit parameters results are reported in \autoref{tab:dwind}, while the best-fit model and residuals are shown in \autoref{fig:spec_dw}. The continuum parameters are only marginally affected by replacing the \xstar table for the \ufoh with the \dwind table, except for a significantly smaller scatter fraction. This is due to the fact that the whole primary continuum is suppressed by the obscuring materials in the wind, and the intrinsic, absorption corrected luminosity is higher: $L_{\rm 2-10\,keV}=5.4$ vs. $1.2\times10^{44} \ergs$. The WA and \ufos parameters are also only marginally affected with respect to the fit with the \xstar tables. 

When fitted as free parameters, the wind terminal velocity and inclination angle are well constrained (see \autoref{fig:cont_dw} left): the \los intercepts the wind at an angle $\theta\sim58^\circ$ from the vertical axis, where the flow has a wide range of velocities and ionization states, which blend the FeXXV and XXVI line pairs into two rather broad absorption features. 
$\fv$ $\sim 1.5$ means that the inner edge of the wind is accelerated from $\vesc=0.25c$ at the origin (R$_{\rm esc}=32\rg$) to $\vinf=0.38$. 

Due to interpolation limitations of the current \dwind grids \citep{Matzeu22a}, it is recommended to freeze these two parameters to the closest grid point ($\mu=0.525$ and $\fv=1.5$, respectively, in this case) in order to get more reliable constraints on the other two parameters of interest $\mw$ and $\lx$. \autoref{fig:cont_dw}, right, shows the results of this approach: the mass outflow rate is broadly consistent with the Eddington accretion rate ($\mw\sim\medd$), while the luminosity seen by the wind is $\lx\sim2.2\%\ledd$, which for a $2\times10^8 \Msun$ BH translates into $\lx=5.5\times10^{44} \ergs$. This is a further indication that the absorption-corrected luminosity in the \dwind fit case is higher
than in the \xstar fit ($L_{\rm 2-10\,keV}=5.4\times10^{44} \ergs$) where Compton scattering is not included: the wind sees a factor $\sim4$ higher intrinsic luminosity with respect to the uncorrected, observed value, as the intrinsic luminosity is affected by obscuration by the thick wind itself.

Applying a bolometric correction of $\kbol\sim50-60$, proper for such luminosity \citep{Marconi04,Duras20}, we derive 
a bolometric luminosity $\lbol\sim3\times10^{46} \ergs$.
This is a factor $\sim2$ larger than the one derived from the IR emission \citep{Veilleux2013},
consistent within the BH mass and bolometric correction factor uncertainties. 

The fit quality is comparable to the fit with three \xstar tables for one more free parameter. In fact, the residuals for the two fits are indistinguishable (see \autoref{fig:broadb} (d) and \autoref{fig:spec_dw}).
We note, however, that neither the \xstar tables nor the \dwind one are able to perfectly model the feature at 11 keV, while they do model the one at 9 keV. This may be related to the fact that the wind is probably more complex in terms of clumpiness, velocity, and ionization structure than the simple geometries assumed in both these approaches.

The \dwind table also includes emission from the outflowing gas outside the \los. This produces a broad emission between 6 and 7 keV rest-frame in a classical P-Cygni profile, with an intensity that depends mainly on the system's geometry. In the best fit described above, this emission partly accounts for the narrow emission modeled with Gaussians at 6 and 7 keV.

We will present in Paper II results on the physical properties of the wind derived using state-of-the-art wind models, such as \wine \citep{Luminari21}, that computes both absorption and emission from the wind in a bi-conical geometry, and \xrade \citep{Matzeu22a}, an emulator of \dwind developed to solve interpolation limitations of this model.

\section{Properties of the X-ray corona}
\label{sec:coronal}

\begin{figure}
\centering
\includegraphics[width=0.45\textwidth]{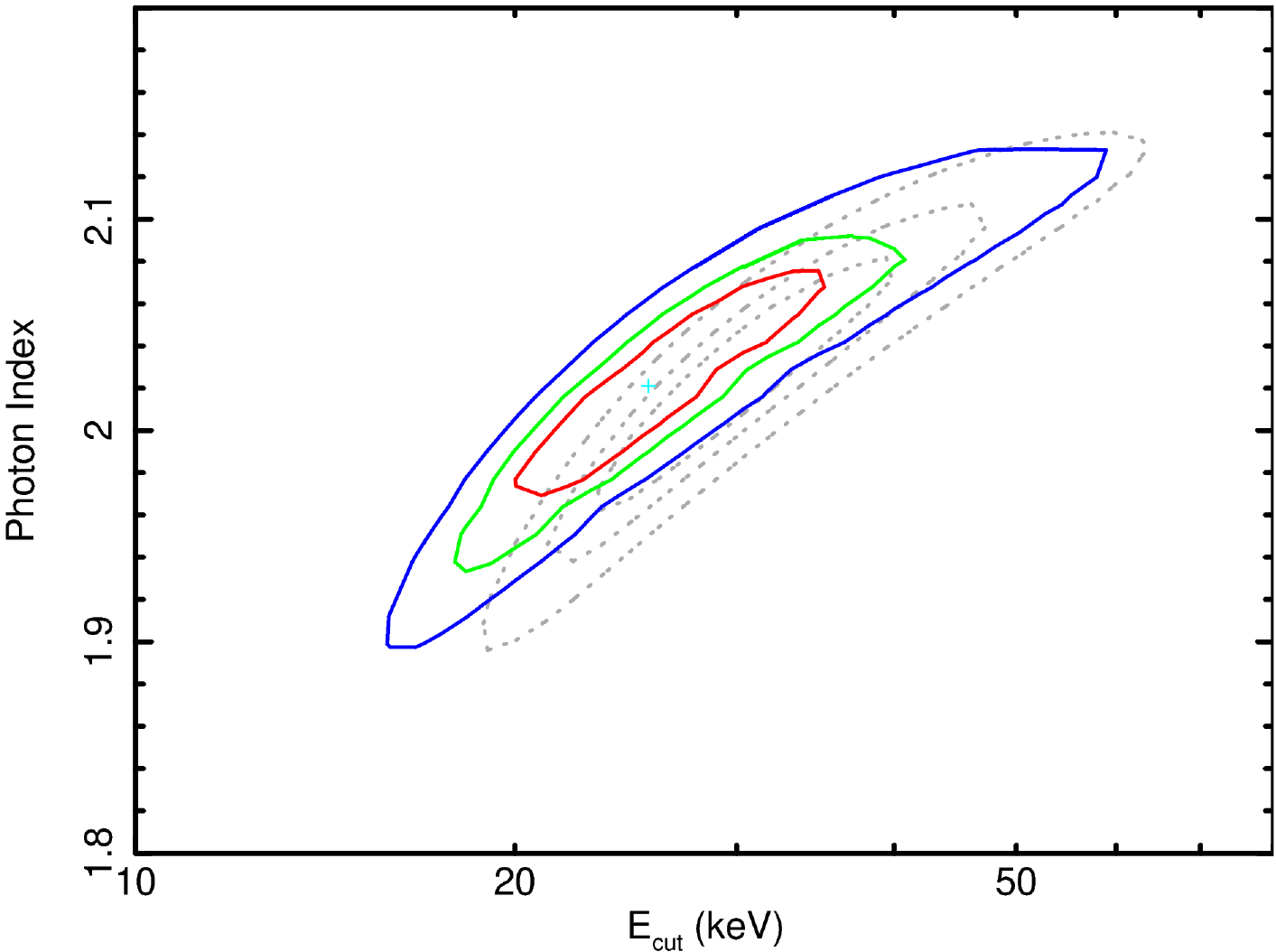}
\caption{Confidence contour levels of the high energy cutoff vs. the photon index. The colored contours refer to the best-fit model with two UFO \xstar absorption tables, described in Sec.~\ref{subsec:xstar}, while the gray contours show the results from the continuum model only, described in Sec.~\ref{subsec:continuum}.}
\label{fig:ecut}
\end{figure}

As described in Sec.~\ref{subsec:continuum}, the primary power-law in IRASF11119 ($\Gamma\sim2$) is modified by an extremely low high-energy cut-off of the order of $\sim30$ keV, plus negligible reflection. By adding the \xstar or \dwind absorption tables, the $\ecut$ value shifts to an even lower value ($\ecut\sim25$ keV, see \autoref{tab:xstar} and \autoref{tab:dwind}), among the lowest reported in the literature \citep{Kara17,Tortosa23}. \autoref{fig:ecut} shows the contour plot of the primary power-law high energy cut-off vs. photon index from the best-fit model with the UFO \xstar absorption tables, described in Sec~\ref{subsec:xstar}, while the gray contours show the results from the continuum model only, described in Sec.~\ref{subsec:continuum}.

\begin{figure}
\centering
\includegraphics[width=0.47\textwidth]{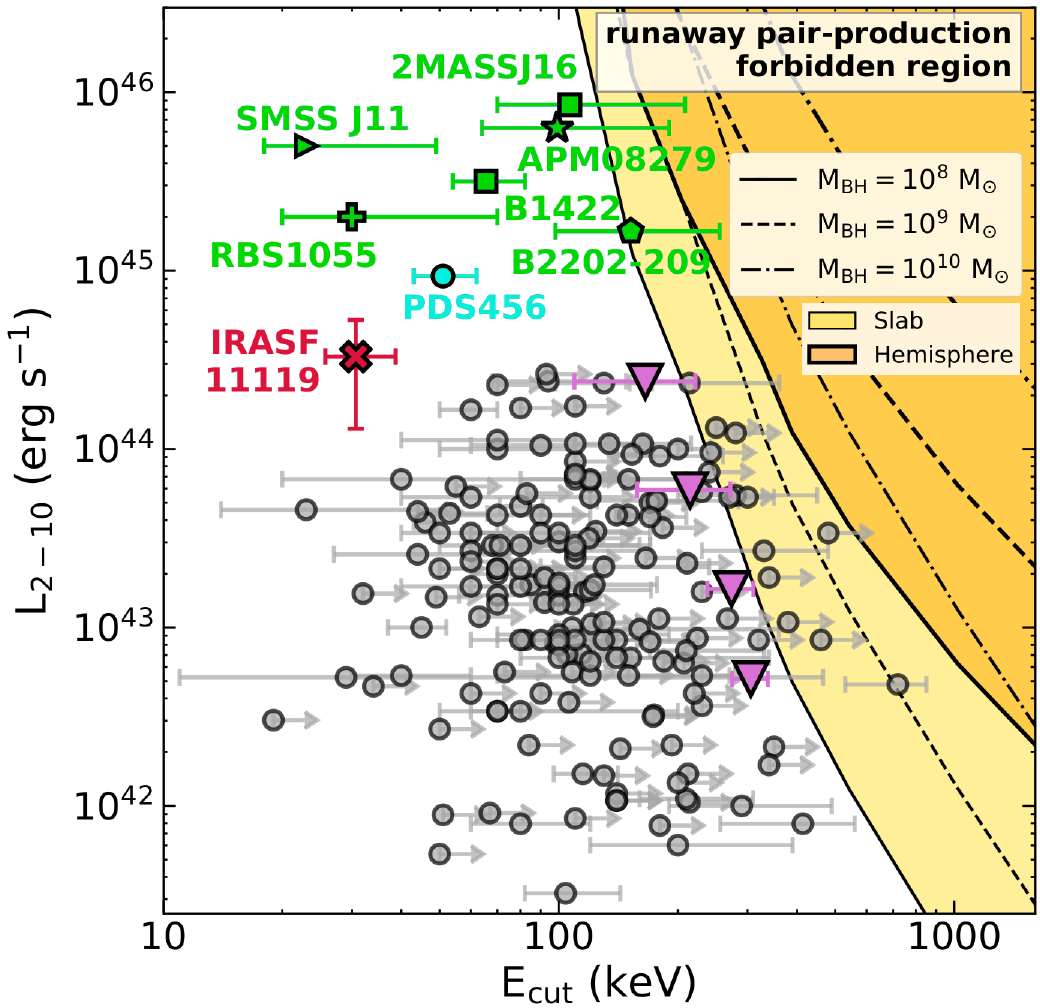}
\caption{High-energy cut-off vs. intrinsic luminosity for local Seyferts measured with \nustar (gray dots, from the updated sample of \citealt{Bertola22}), and the binned results from the BAT-selected sample (magenta down-ward triangles, \citealt{Ricci18MNRAS.480.1819R}). Green data points show results for luminous quasars at $z\gtrsim0.5$: pentagon from \cite{Kammoun17}; squares \cite{Lanzuisi19}; star \cite{Bertola22}; plus \cite{Marinucci22}; right-ward triangle \cite{Kammoun23}.
PDS456 \citep{Reeves21} is shown in cyan and IRASF11119 in red. The different intrinsic luminosities obtained from the \xstar and \dwind fits are shown by the vertical error bar.
The yellow/orange areas show the runaway pair-production forbidden region \citep{Fabian15,Lanzuisi19} for different geometries and BH masses.
}
\label{fig:ecut_lum}
\end{figure}

We note that the high energy cut-off and the photon index of the primary power-law modeled with \pexrav are correlated, as a flatter photon index requires an even lower cut-off to account for the observed lack of hard band counts. Also, the addition of absorption tables, such as \xstar or \dwind, shifts the $\ecut$ to lower values, as part of the soft emission is absorbed by the wind, and therefore, the continuum in the hard band can be even steeper. However, the results are consistent with each other within $1\sigma$ confidence level.

Thanks to the advent of the \nustar hard X-ray imaging telescope \citep{Harrison13}, the high energy cut-off is now known with unprecedented accuracy for tens of nearby bright Seyfert galaxies \citep[e.g.][]{Fabian15, Tortosa18a, Kamraj18, Balokovic20, Kang22}. 
\autoref{fig:ecut_lum} shows these measurements as a function of X-ray luminosity for local Seyferts (gray circles, from the updated sample in \citealt{Bertola22}) and the binned results from the BAT-selected sample of \citealt{Ricci18MNRAS.480.1819R} (magenta triangles). The few luminous quasars at $z\gtrsim0.5$ 
for which such measurement has been possible so far are shown in green \citep{Kammoun17, Lanzuisi19, Bertola22, Marinucci22, Kammoun23}, while PDS456 \citep{Reeves21} is shown in cyan. The yellow/orange areas show the runaway pair-production region \citep{Fabian15}, translated into directly observable quantities for different geometries of the corona and assuming different BH mass values, shown with different line styles (see \citealt{Lanzuisi19} and \citealt{Bertola22} for details).
The $\ecut$ measured for IRASF11119 (see \autoref{fig:ecut}) is shown by the red cross, where the luminosity is the average of the two different measurements obtained from the \xstar and the \dwind fits, and the error bars show the range covered by them. The target shows one of the lowest $\ecut$ reported in the literature at all luminosities.

Several examples are known of strong X-ray UFOs detected in sources with relatively cool coronae, such as PDS456 \citep[$\ecut\sim50$ keV,][]{Reeves21}, or LBQS~1338-0038 ($\ecut\sim50$ keV; \citealt{Matzeu23}, Matzeu et al. in prep.).
The super-Eddington accreting NLSy1s I~Zw~1 \citep{Ding22ApJ...931...77D} and IRAS~04416+1215 \citep{Tortosa22} show even lower $\ecut$ values ($\sim15$ keV) and are again associated with the presence of UFOs. 
These findings indicate a possible correlation between strong nuclear outflows and a cold corona or soft continuum, often associated with highly accreting/super Eddington and NLSy1 sources. 

High Eddington rates, by definition, promote strong radiation-driven winds and, at the same time, increase the efficiency of the corona cooling, as the Compton cooling is proportional to the inverse of the radiation energy density.
In fact an anti-correlation between $\ecut$ and $\eddratio$ has been indeed observed in \citet{Ricci18MNRAS.480.1819R} based on BAT data, even if such anti-correlation has not been confirmed in later studies based on  \nustar data \citep{Hinkle21,Kamraj22ApJ...927...42K,Serafinelli23}.
In a systematic study of super-Eddington sources \cite{Tortosa23} indeed found indication of softer spectra and the presence of ionized outflows in most of them, but only two out of eight show low $\ecut$ values.

On the other hand, a colder corona or softer power-law may promote the formation of nuclear winds by preventing the gas surrounding the inner disk from being over-ionized \citep{Proga04, Nomura20, Matzeu22a}, 

In particular, in case of a high accretion rate, 
it has been suggested \citep[e.g.][]{Sadowski16,Kubota19} that the inner standard accretion disk puffs up due to the high radiation pressure of UV photons.
The inner puffed-up disk increases the UV/soft X-ray flux seen by the corona, leading to more efficient cooling and at the same time, shields
the rest of the inner accretion disk from hard X-rays, preventing over-ionization and allowing 
for inner/faster winds to be launched.
This mechanism has been invoked for z>6 QSOs showing a correlation between photon index and nuclear wind (observed as CIV emission line outflowing components) outflow velocity \citep{Tortosa24sub}.

The connection between a steep continuum and the presence of strong/persistent nuclear outflows may also have important implications for the impact of AGN feedback in the early Universe, given the recent results on the spectral steepening in the $z>6$ QSO HYPERION sample \citep{Zappacosta23}.
A systematic study of the incidence of UFO features as a function of coronal properties and the Eddington ratio is thus needed to explore the link between all these aspects of SMBH accretion, but is beyond the scope of this paper.

\section{Large-scale wind properties from optical spectroscopy}
\label{sec:ionized}

\begin{figure*}[t]
    \centering
    \includegraphics[width=\textwidth]{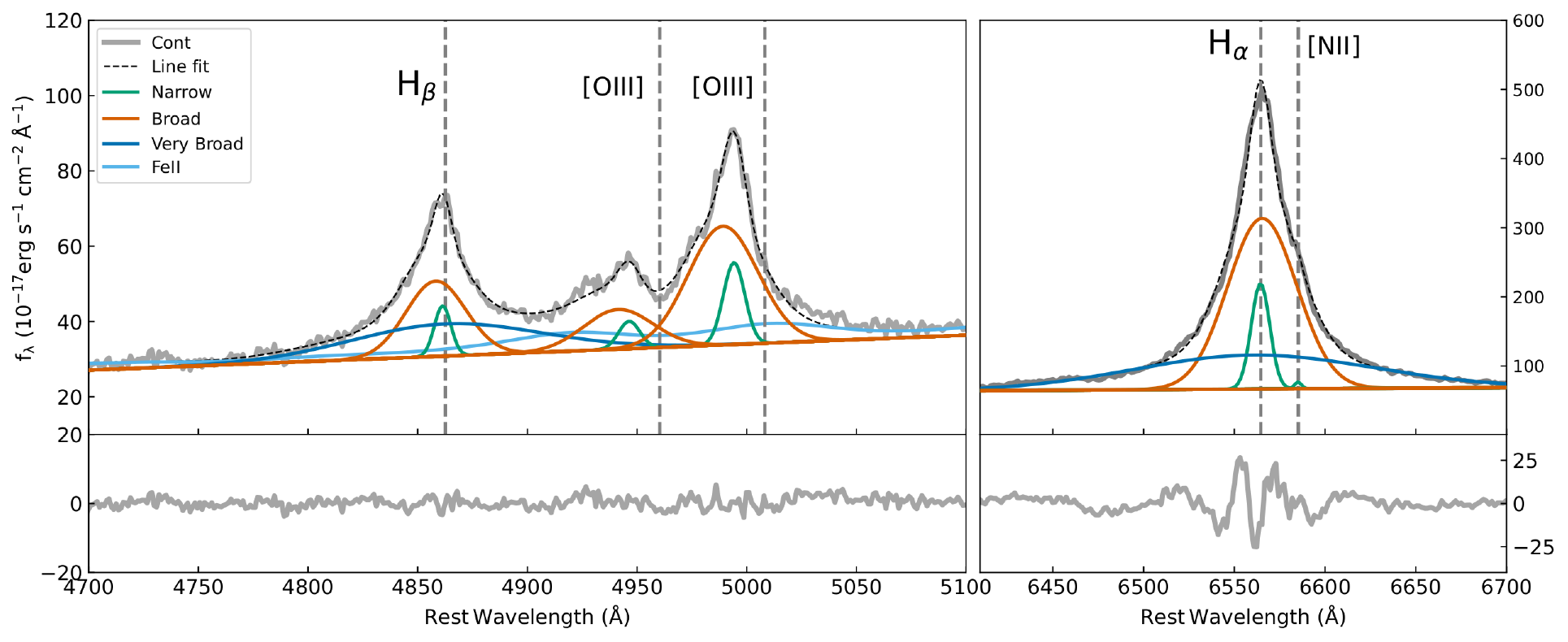}
    \caption{H$_\beta$, [OIII] and H$_\alpha$ regions of the optical spectrum of IRASF11119. The observed spectrum is plotted in grey, while the best fit is shown as the black dashed line. Different model components are traced by lines of different colors, as indicated by the label. The vertical dashed lines indicate the rest-frame wavelength for each labeled emission line, for z=0.187. In the bottom panels, residuals are plotted. Note that the H$_\alpha$ shows significant residuals, possibly due to the blending with the [NII] doublet, but this effect is negligible for our goals.}
    \label{fig:optfit}
\end{figure*}

The optical spectrum of IRASF11119 shows peculiar features that have been only partially analyzed in the literature \citep{Lipari03,Rupke05}. We reanalyzed the publicly available 2013 SDSS-eBOSS spectrum of IRASF11119 by performing simultaneous multi-component fitting using the code PyQSOFit \citep{guo2018pyqsofit}.
We model the continuum with a polynomial component and FeII emission templates from \citet{boroson1992emission}.
Galactic extinction is also included through the dust-reddening maps of \citet{schlegel1998maps}.
We simultaneously model emission lines with a combination of narrow and broad Gaussian functions. These include:
\begin{itemize}
    \item \textbf{Six} narrow (FWHM $<$ 500 km/s) lines:  
    H$_\alpha$ (6563 $\mathrm{\AA}$), H$_\beta$ (4861 $\mathrm{\AA}$), and [OIII] (4959,5007 $\mathrm{\AA}$) and [NII] (6549,6584 $\mathrm{\AA}$) doublets. The ratios of the [OIII] and [NII] doublets were both set to be 1:2.99 from atomic physics.
    \item \textbf{Two} very broad (FWHM $>$ 2000 km/s) lines to account for the H$_\alpha$ and H$_\beta$ Broad Line Region (BLR) emission.
    \item \textbf{Six} broad (500 km/s $<$ FWHM $<$ 2000 km/s) lines to account for outflows or to better describe the BLR emission profile, which is known to be poorly reproduced with only one Gaussian line. The ratios of the [OIII] and [NII] doublets were constrained as above.
\end{itemize}

We constrain the width of the narrow components to be the same, and we do so for the broad components as well. Initially, we also request for the line centroid offset to be the same for all the lines belonging to the same group (narrow, broad, and very broad), as it is a well-tested recipe used in the characterization of galactic outflows \citep[e.g.][]{brusa2015x,perna2015galaxy}. However, these constraints appear to be too harsh to model the complex kinematics revealed by the optical spectrum, and we, therefore leave the offset free to vary from one line to the other in the same group.

\begin{figure}
\centering
\includegraphics[width=0.47\textwidth]{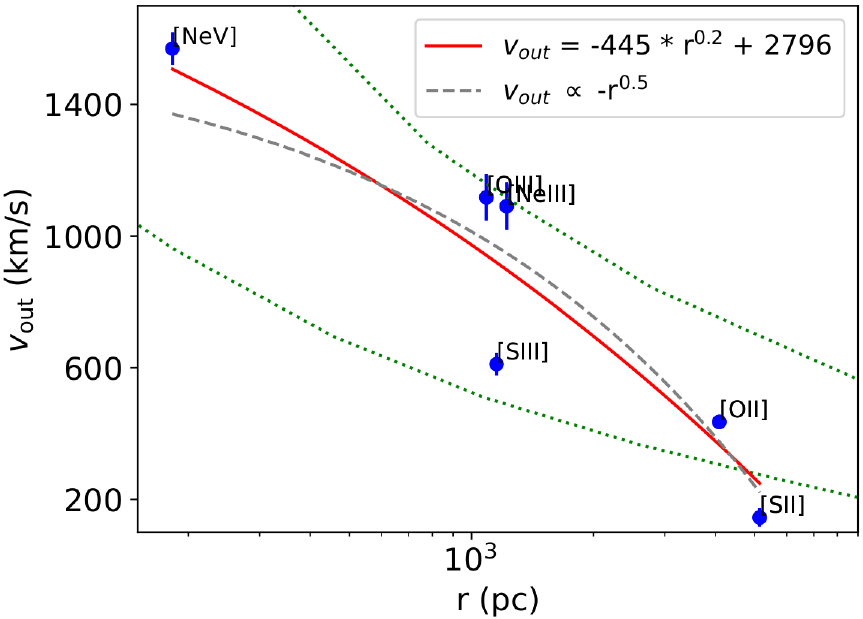}
    \caption{Distance from the ionizing source vs. velocity offset for different emission lines (labeled) modeled in the SDSS spectrum. The distance is derived from the ionization level at which each ion has the peak of relative abundance.
    The red curve is a power-law fit of the form $v_{out} = a\times r^b+c$ with the best-fit slope $b=0.20\pm0.07$. The gray dashed line shows a curve with a slope fixed to $b=0.5$, the value for a simple ballistic trajectory. The dotted green curves are taken from the analytic treatment of an expanding AGN-driven wind presented in \cite{Fauchere12}, for $v_{in}=0.1c$, and central density $n_0=10$ (top) and $n_0=100$ (bottom) $\cmq$.}  
    \label{fig:lines}
\end{figure} 

IRASF11119, in fact, is a "blue outlier", meaning that the peak of the [OIII] emission lines is blue-shifted by more than 250 km/s with respect to the Balmer lines \citep{zamanov2002kinematic}. These rare sources (3-5\% of optically selected AGN) are typically associated with low-mass/highly-accreting AGN and with the presence of nuclear outflows \citep{Marziani03,Marziani16}. Indeed, while the H$_\beta$ and H$_\alpha$ peak at wavelengths corresponding to z=0.190, the [OIII] lines peak corresponds to z=0.187, with little to no narrow emission at the expected wavelength for z=0.190. The whole [OIII] emission can, therefore be interpreted as due to outflowing material. We decided to untie the centroid offset of the lines in the [OIII] doublet from the rest of the narrow lines while keeping them tied to each other. In this way, both narrow and broad [OIII] lines account for outflow emission. The fit results are shown in Fig. \ref{fig:optfit}. 

For the BLR, the FWHM of the broad and very broad components combined is $\sim2500$ km/s, comparable with the $2000$~km/s adopted in \citet{NardiniZubovas} from which they derived a black hole mass of $\mbh\sim2\times10^8$ \Msun.

In addition to the previously reported lines,
we also detect emission from forbidden lines of different ionization levels such as [SII], [SIII], [OII], [NeIII], and [NeV].
For each line, we measure a non-negligible  
velocity shift with respect to the systemic redshift obtained from the Balmer lines, and we found a clear positive trend between this shift and the ionization level of the gas responsible for each line.
A similar trend is usually observed between the [OIII] and the Balmer lines \citep{Veilleux91a,Zamanov02,Eracleous04,Hu08}. 
This is interpreted in the context of decelerating outflows in a stratified medium, photoionized by the AGN \citep{Veilleux91a,Zamanov02,Komossa08,Spoon09}: 
More highly ionized species correspond to gas closer to the nuclear AGN, and the observed trend indicates a stratification of the gas velocity, decreasing at larger radii.

To test this hypothesis, we computed the ionization parameter at which each ion has its peak of relative abundance using \cloudy, v23.0 \citep{Cloudy23}, assuming a fixed gas density of $n_H=10^3 \cmq$ and the SED of IRASF11119 as the ionizing continuum. We can then convert this into a distance from the ionizing source using the definition of the ionization parameter $\displaystyle\xi \equiv {L_{ion}}/{nr^2}$ where $n$ is the number density of the absorbing material, and $r$ is the distance.

The results are shown in \autoref{fig:lines}. 
Given the low S/N levels of the SDSS spectrum, the [NeIII], [OII], and [SII] doublets are fitted with pairs of lines at a distance fixed to the lab value so that we derive a single velocity offset value for each doublet. With our assumptions, we derive that the high-ionization line [NeV] is produced by gas at $\sim200$ pc from the center, outflowing at velocities of $\sim1500$ km/s, while [OII] and [SII] doublets are produced by gas at few kpcs (i.e. at scales comparable with the whole host galaxy) and at that point the outflow velocity has dropped to $\sim200$ km/s. 
The red curve is a power-law fit to the data in the form $v_{out} = a\times r^b+c$. The best-fit slope is $b=0.20\pm0.07$. The dashed gray line shows a dependency of the form $r^{0.5}$, i.e., a purely ballistic trajectory. The best-fit relation is slightly flatter but consistent with the ballistic curve.
AGN radiatively-driven large-scale winds are expected to be energy-conserving at these scales and to propagate with a flatter velocity gradient, thanks to the pressure support of the hot wind \citep[see, e.g.,][]{Fabian15}. This is shown by the green dotted curves taken from the analytic treatment presented in \cite{Fauchere12}, with parameters $\lbol=10^{46} \ergs$, $v_{in}=0.1c$, and central density $n_0=10-100 \cmq$. 

We stress that different assumptions on the value of the constant gas density can substantially change the results in terms of the radius of each ion, while the general trend of decreasing velocity at larger radii would remain.
Introducing a radial-dependent density profile would change the slope of the curve, i.e., flatter curves for density decreasing with radius.
A more detailed investigation taking into account the possible effects of a clumpy medium and of the different line critical densities is beyond the scope of this paper.

\section{Energetics of the outflow components}
\label{sec:ene}

\subsection{Nuclear Outflow}
To derive the energetics of the nuclear outflow we followed a standard approach starting from the best-fit values of the \xstar tables described in Sec.~\ref{subsec:xstar}. 
A lower limit of the distance between the central BH and the absorbing gas layer can be inferred from the radius at which the observed outflow velocity along the line of sight corresponds to the escape velocity:
\begin{equation}
    r_{min} \equiv \frac{2 G M_{BH}}{v_{out}^2}
\end{equation}

For the derivation of the maximum distance of the absorber, the definition of the ionizing luminosity L$_{ion}$ is often used, which is the source ionizing luminosity integrated between 1-1000 Ryd. We computed L$_{ion}$ from the broad-band SED derived in Sec.~\ref{sec:data_red}, i.e. $\lion=4\times10^{45} \ergs$.
We use again the definition of the ionization parameter, $\displaystyle\xi \equiv {L_{ion}}/{nr^2}$. As the size of the absorber cannot exceed its distance to the black hole, $N_H \simeq n \Delta r < nr$, we obtain:
\begin{equation}
    r_{max} \equiv \frac{L_{ion}}{\xi N_H}
\end{equation}
In the case of IRASF11119, $r_{min}\sim13R_S$ and $r_{max}\sim530R_S$ for the \ufoh absorber. Following the same approach for the \ufol and WA absorbers would translate into $r_{min}\sim11R_S$ and $r_{max}\sim10^7R_S$ or $\sim650$pc for \ufol and 
$r_{min}\sim10^5R_S$ or a few pc, and $r_{max}\sim10^9R_S$ or $\sim10^4$ pc for the WA. 
These simplified definitions are, therefore, clearly not useful for these lower ionization absorbers since the values of $r_{min}$ and $r_{max}$ span several orders of magnitudes from the vicinity of the SMBH to the outskirts of the host galaxy. 

In fact, estimates of the location of the WA vary significantly in the literature, possibly reflecting the intrinsic multi-phase, multi-temperature nature of these outflows \citep[e.g.][]{Krolik01,Krongold07,Digesu13,Laha21NatAs}.
Studies from extensive multi-wavelength campaigns on single sources and based on variability \citep[e.g.][]{Kaastra11} locate the WA at tens of pc from the SMBH \citep{Kaastra12,Kaastra14}.
The location of the gas responsible for the \ufol is even more uncertain. 
It has been suggested, and in at least one case derived from variability arguments \citep{Serafinelli19}, that these 
low column and ionization UFOs may be located at the interface between the nuclear UFO and the WA, i.e., hundreds of parsecs from the SMBH \citep[e.g.,][]{Gaspari20}, while 
the interpretation as the UFO gas cooling down and fragmenting \citep[e.g.,][]{Takeuchi13}  implies much smaller distances (hundreds of $\rg$).

In any case, it is to be expected that the energy carried by both the \ufol and the WA is only a small fraction of that of the "classical" UFO, as the column density is 2-3 orders of magnitude lower, and for WA the outflow velocity is negligible (see \citealt{Reeves18PG1211} for a notable exception). Therefore, in the following, we focus on the derivation of the energetics of the \ufoh component only. 

We compute the mass outflow rate through the standard expression: 
\begin{equation}
\dot{M}_{out} \simeq m_{\rm p} \Omega \nh \vout r,
\label{e0}	
\end{equation}
following \citet{Tombesi12, Gofford13, NardiniZubovas} and assuming solar abundances, full ionization, and neglecting more complex geometrical descriptions. $m_{\rm p}$ is the proton mass, $\Omega$ is the solid angle subtended by the wind, and $\nh$, $\vout$ and $r$ are its column density, velocity, and radius. 
We adopt a solid angle $\Omega_{\rm FeK}/4\pi = 0.5$ consistent with the value derived in Sec.~\ref{subsec:xstar}, and comparable to the one derived from the P-Cygni profile in PDS 456 \citep{Nardini15} and from population studies \citep{Tombesi10,Gofford13}.

We compute $\dot{M}_{out}$ using the lower limit reported in \autoref{tab:xstar} as the upper boundary is unconstrained. The measured $\vout$ is also a lower limit since the observed value is only the component along the l.o.s. 
Consequently, we compute $\dot{M}_{out}$ only for the $r_{min}$ as we can only derive a lower limit of this quantity. 
We further note that taking into account special relativistic effects, the observed $\nh$ must be corrected to take into account the reduction of the optical depth of the outflowing gas for increasing $\vout$ \citep[e.g.][]{Luminari20}.
This velocity-dependent correction is of the order of $\sim2$ for an outflow velocity of 0.27c. 
In the following, we take into account this correction when 
computing the wind energetics.
We therefore obtain $\mout>2.10 \Msun/yr$ or $\mout\gtrsim0.49 \medd$ or $\mout\gtrsim0.82 \macc$. 

In the classical treatment, we can then derive the mechanical power as $\dot{E}_{kin} \equiv \frac{1}{2} \, \dot{M}_{out} \, v_{out}^2$ 
and the outflow momentum rate
$    \dot{P}_{out} \equiv \dot{M}_{out} \, v_{out}  $.
The correction on $\dot{E}_{kin}$ due to the special relativistic formula $(\gamma-1) \mout c^2$ \citep{Saez11,Luminari21} is minor since the Lorentz factor $\gamma$ is $\sim1.04$ for $\vout=0.27c$.
The results for {UFO$_{\rm H}$ are therefore 
$\dot{E}_{kin}>0.32 \lbol$ and $\dot{P}_{out}>2.26 \lbol/c$.

In the \dwind fit instead, it is possible to directly derive the mass outflow rate and the terminal velocity of the wind, corrected for projection effects, as $\mw$, $\fv$, and $\mu$ are parameters of the fit. 
We use the $\mout$ derived in \autoref{tab:dwind}, 
$\mout=0.99_{-0.17}^{+0.26}~\medd$, that for a $\mbh=2\times10^8 \Msun$ translates into $4.25_{-0.73}^{+1.11} \Msun/yr$ or $1.64_{-0.28}^{+0.43}\macc$ given the observed mass accretion rate. 

Therefore, in IRASF11119, as it happens in similar highly accreting sources (PDS456, PG~1211+143), the mass outflow rate is comparable with the mass accretion rate and close to the Eddington limit. This suggests that the accretion rate in the external part of the accretion disk is potentially higher than Eddington, and the outflow substantially contributes to keeping the effective accretion rate onto the SMBH below such a limit. 

The derived terminal velocity is larger than the observed one obtained from the \xstar fit. This is due to inclination (deprojection) effects and to the acceleration of the gas along the line of sight up to the maximum velocity. Also, this is the velocity corresponding to the inner edge of the wind (i.e., R$_{\rm esc}=32\rg$), while the outer edge (i.e., R$_{\rm esc}=48\rg$) has a lower terminal velocity of 0.3c.
The mechanical power and outflow momentum rate are also higher than what was estimated from the \xstar fit:
$\dot{E}_{kin}=1.21_{-0.20}^{+0.32}\ \lbol$ and $\dot{P}_{out}=6.37_{-1.09}^{+1.67}\ \lbol/c$.

\subsection{Large-scale ionized outflow}

We analyze the kinematic of the large-scale outflow observed in the [OIII] emission in the SDSS spectrum using prescriptions well established in the literature (e.g., \citealt{cano2012observational}, and \citealt{Fiore17}).
By assuming spherical geometry and constant density of the outflow, the mass outflow rate can be computed as $\mathrm{\dot{M}_{OF}} = 3 \times \mathrm{V_{OUT}} \times \mathrm{M_{OF}} \times \mathrm{R_{OF}^{-1}}$ (see \citealt{Lutz20} for a discussion on the implication of different assumptions on outflow geometry and history).
$\mathrm{V_{OUT}}$ is the maximum outflow velocity, $\mathrm{R_{OF}}$ is the outflow radius and $\mathrm{M_{OF}}$ is the mass of the gas entrained in the outflow.
We constrain $\mathrm{R_{OF}}$ using the SDSS-BOSS fiber diameter (2"), which corresponds to 3.2 kpc at z=0.190. This should be considered as an upper limit, and it is roughly consistent 
with  what is derived in \autoref{fig:lines} for the [OIII] radial extent.
Therefore, $\mathrm{R_{OF}}$ and $\mathrm{M_{OF}}$ should also be considered upper limits.

The outflow velocity $\mathrm{V_{OUT}}$ is estimated through non-parametric analysis in order to minimize bias arising from the fitting procedure. We compute both $v_{02}$ and $w_{80}$. $v_{02}$ is the velocity corresponding to 2\% of the flux for the [OIII]$_{\mathrm{\lambda5008}}$ line, with respect to the rest frame emission. $w_{80}$ corresponds to the difference in velocity between 10\% and 90\% of the flux of the [OIII]$_{\mathrm{\lambda5008}}$ line. The second velocity estimator, $w_{80}$, is thought to be more robust, as it takes into account a difference between velocities rather than just one measurement. However, it is not sensitive to the fact that \textit{the whole} [OIII] emission is blue-shifted. Our estimate of $\mathrm{V_{OUT}}$ is therefore $v_{02}= 3053 \pm 92$ km/s, rather than $w_{80} = 2264 \pm 68$ km/s. 

The mass $\mathrm{M_{OF}}$ is given by:
\begin{equation}
    \mathrm{M_{OF}} = 5.3 \times 10^7 \frac{L_{44}(\mathrm{[OIII])}}{n_{e_3}10^{[\mathrm{O/H}]}} \ \mathrm{M_{\odot}}
\end{equation}

Where $L_{44}(\mathrm{[OIII]})$ is the luminosity of the [OIII]$_{\mathrm{\lambda5008}}$ line tracing the outflow in units of $10^{44}$ erg/s, $10^{[\mathrm{O/H}]}$ is the metallicity of the outflowing gas and $n_{e_3}$ is the electron density of the same gas in units of 1000 cm$^{-3}$.
Most authors estimate $n_{e_3}$ through the ratio $R$ between the lines of [SII] doublet. However, in our case, the low SNR did not allow for robust measurement of $R$. In addition, this method has been shown to be biased towards lower densities, also in samples of AGNs with outflows (e.g., \citealt{baron2019discovering,davies2020ionized}).
For these reasons, we conservatively assume the median density $log(n_{e}) = 3.7 \pm 0.16 $ derived in \citet{rose2018quantifying} using the \textit{Trans-Auroral Ratio} (TR) which measures $n_{e_3}$ making use of the doublets [SII]$_{\lambda \lambda \ 4068,4076}$, [SII]$_{\lambda \lambda \ 6717,6731}$, [OII]$_{\lambda \lambda \ 3726,3729}$, and [OII]$_{\lambda \lambda \ 7319,7331}$. 

The observed [OIII]$_{\mathrm{\lambda5008}}$ luminosity is $L(\mathrm{[OIII]})=1.59 \pm 0.11 \times10^{42}$ erg/s. However, the spectrum is reddened, as indicated by the Balmer decrement H$_{\alpha}$/H$_{\beta}$ = $12.6 \pm 2.9$, which is $>3-4$ times higher than the standard intrinsic value of 2.86 (\cite{1989agna.book.....O}). By using the empirical approach presented in \citet{calzetti1994dust}, and using the extinction curve $k(\lambda)$ found in \citet{calzetti2000dust}, we derive the corrected value of $L(\mathrm{[OIII]})=1.61^{+1.46}_{-0.92} \times10^{44}$ erg/s. Assuming solar metallicity, we derive $\mathrm{\dot{M}_{OF}} = 49^{+45}_{-28} \ \mathrm{M_{\odot}/yr}$. 
This corresponds to a momentum outflow rate of 
$\mathrm{\dot{P}_{OF}} = 1.90^{+1.98}_{-1.15} \ L_{bol}/c$.

The energetics recently derived in \citet{pan2019discovery} 
using the same SDSS spectrum, along with dedicated NIR-MIR data, 
is higher by a factor $\sim7$ of the one derived above.
This is partly due to the different assumed geometries: while we assume a uniformly filled spherical geometry, they assume co-spatiality and coupling of molecular, neutral, and ionized gas in a thin-shell geometry, which corresponds to higher mass outflow rates (e.g., \citealt{maiolino2012evidence}).
More importantly, their modeling of the outflow combines information from molecular, ionized, and neutral phases of the gas. Consequently, their derived parameters are similar to those derived in \citet{nardini2018multi} for the molecular outflow alone, which is dominating the energetics at large radii. Our method, instead, by focusing on [OIII], probes the faster, lighter, and more highly ionized phase of the large-scale outflow.

\subsection{Momentum boost diagram}

\begin{figure}
\centering
\includegraphics[width=0.47\textwidth]{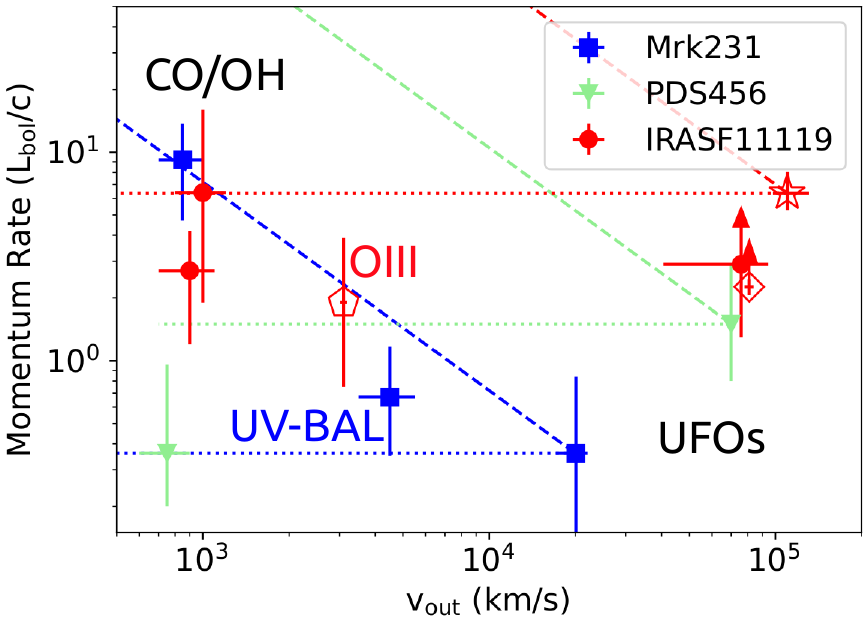}
\caption{Momentum rate vs. outflow velocity of different outflow phases for three well-known targets. Red symbols refer to IRASF11119: filled circles are taken from \cite{NardiniZubovas}. The empty diamond shows the \ufoh momentum rate derived from the \xstar fit, while the empty star shows the results from the \dwind fit. 
The empty pentagon shows the momentum rate of the large-scale ionized outflow observed in [OIII].
Blue squares refer to Mrk231 \citep{Feruglio15} and green triangles to PDS456 \citep{Bischetti19PDS}.
Data points at $\vout\sim10^3 \kms $ refer to molecular outflows detected in CO or OH. Data points at intermediate velocities ($3-5000 \kms$) refer to OIII and UV winds, while at $\vout>10^4 \kms$ refer to UFOs. 
Dotted horizontal lines show the expected evolution of a momentum-driven outflow, while dashed lines show the momentum boost 
expected for an energy-driven outflow.
}
\label{fig:ene}
\end{figure}

\autoref{fig:ene} summarises the results on the wind energetics derived for IRASF11119 (in red). The diamond shows the \ufoh momentum rate derived from the \xstar best-fit parameters, while the empty star shows the results from the \dwind model fit. The empty pentagon shows the momentum rate of the large-scale ionized outflow observed in [OIII]. The filled circles show values from the literature, as recomputed in  \citet{NardiniZubovas}.
We note that the error bars on $\vout$ from the \xstar fit presented in \autoref{fig:ene} are much smaller than those used in \cite{NardiniZubovas} and derived from the \nustar follow-up of 2016 \citep{Tombesi17}, thanks to the much higher spectral quality and energy resolution of our data. On the other hand, the error bars on $\vinf$ from the \dwind fit are larger and more realistic than those obtained from the \xstar fit, as the \dwind fit is taking into account the systematic uncertainties related to the system's geometry.

We compare our target with two local QSOs/ULIRGS systems, Mrk231 and PDS456, chosen as two of the best-studied multi-phase outflows and at the two extremes of the wind propagation mechanisms: PDS456 also known as "the gentle monster" \citep{Bischetti19PDS}, has a large scale outflow detected in CO(3-2) with ALMA, with a momentum flux much smaller than the one measured from the very well studied and persistent X-ray UFO \citep[e.g.][]{Nardini15, Matzeu16}. Mrk231, on the other hand, exhibits a powerful kpc-scale, multi-phase (molecular and highly ionized, UV-BAL) outflow consistent with an energy-driven expansion powered by the nuclear wind observed in the X-rays \citep{Feruglio15}. 

These two sources are at the two extremes of the broad range of momentum boosts observed in the increasing sample of sources from which the energetics of the different outflow phases have been derived. In general, they show behaviors following either the momentum- or the energy-driven framework, with a majority in the first group \citep{Tozzi21,bonanomi23}. 

IRASF11119 is one of the very few sources for which the energetics of the large-scale outflow have been derived both in the molecular and ionized phases.
According to the values derived in this work from the analysis of the new X-ray observations, the momentum flux of the nuclear UFO is very high in terms of $\lbol/c$, 
$\dot{P}_{out}\gtrsim1 \lbol/c$ from the \xstar fit or $\dot{P}_{out}\sim6 \lbol/c$ from the \dwind results, while the momentum fluxes of the large-scale outflows are comparable with what observed in Mrk231 ($\dot{P}_{out}=1-10 \lbol/c$). 
The source, therefore, falls in the predominant group of sources (see \citealt{Smith19, bonanomi23}) for which the large-scale wind appears to be not significantly boosted with respect to the nuclear outflow. 

This could be due to the fact that the flow remains momentum-driven even at large distances from the SMBH, despite the expectations that beyond a few 100s pc the Compton cooling should become ineffective \citep[e.g.][]{King11}, or could be due to an inefficient coupling of the UFO with the host-galaxy ISM \citep{Tombesi15}.
Alternatively, the UFO observed today may be more powerful than it has been - on average - in the past few Myrs (flow time of the CO large scale outflow) so that the large scale outflow is the result of several short bursts of Eddington limited accretion episodes (see \citealt{NardiniZubovas} and \citealt{ZubovasNardini}).
This is consistent with a chaotic feeding mechanism, as predicted by the CCA scenario \citep[e.g.;][]{Gaspari17_cca}.
Unfortunately, we do not have direct access to what has been the duty cycle of the wind over time scales of $10^5-10^6$ years, i.e., the flow time of the OH and CO, respectively, as estimated from the ratio $R_{out}/V_{out}$ in \cite{Veilleux17}.

On the other hand, a purely energy-driven expansion powered by a steady nuclear wind like the one observed in this work would imply large-scale momentum rates of hundreds to thousands times $\lbol/c$, i.e., mass outflow rates of more than several thousand $\Msun/yr$. Such extreme mass outflow rates have been observed only in the most powerful high-z systems, with typical $\lbol>10^{47} \ergs $ \citep{Carniani15A&A...580A.102C,Bischetti17A&A...598A.122B} and in only a couple of extreme radio galaxies \citep{Nesvadba08A&A...491..407N} where the AGN $\lbol$ is comparable to IRASF11119 (few$\times10^{46} \ergs$), but the outflow is most probably powered by the strong radio jets.

\section{Summary}
\label{sec:conc}

In this work, we have analyzed the new \xmm and \nustar data collected for the highly accreting, post-merger ULIRG/QSO system IRASF11119, together with its SDSS optical spectrum. The results can be summarized as follows:

\begin{itemize}
    \item A "classical" nuclear wind with high column and ionization (\ufoh), outflowing at $\vout\sim0.27c$ is clearly detected in the hard band, with parameters comparable to the previous \suzaku and \nustar observations, while a new 
    component with lower column and ionization (\ufol) but similar speed is detected in the soft band. 
    \item We robustly measure one of the lowest high-energy cut-offs observed in AGN, $\ecut\sim25-30$ keV. 
    Several other examples of sources with similar properties suggest that steep X-ray spectra - either soft power-laws or low cut-off - may promote or be connected with strong nuclear winds.
    \item The optical spectrum shows a strong ionized large-scale outflow in the [OIII] forbidden lines, while several other forbidden lines are detected with decreasing outflow velocity for decreasing ionization potential, interpreted as an outflow decelerating, via entrainment, in a stratified medium, photoionized by the AGN.
    \item The energetics derived for the \ufoh, both from standard assumptions applied to the \xstar fit and more directly from the \dwind fit, are large, such that the mass outflow rate is comparable with the mass accretion rate and to the Eddington limit mass, $\mout\sim\macc\sim\medd$.  
    \item The absence of a significant momentum boost between the nuclear UFO and the different phases of the large-scale outflow suggests either a momentum-driven expansion of the wind, or a small coupling between the nuclear outflow and the ISM, or strong variability of the nuclear outflow power on time scales comparable with the large scale outflow flow-time ($\sim0.1-1Myr$).  
\end{itemize}

Including IRASF11119 in the broader context of the SUBWAYS sample of $z=0.1-0.5$ QSOs with comparable X-ray spectral quality gives a total UFO detection fraction of $0.35_{-0.14}^{+0.17}$ (8 over 23 sources), slightly larger than the $\sim30\%$ derived in \cite{Matzeu23} and identical to the value derived in the local Universe in \cite{Tombesi10}.
This is without considering the well-known, persistent UFO detected in PDS 456, which would belong to the $z=0.1-0.5$ sample but is an outlier in terms of BH mass and luminosity.

IRASF11119 will be a natural target for XRISM \citep{Xrism} observations: the high-resolution spectra of Resolve \citep{Resolve18} will allow to i) characterize the physical properties of the soft X-ray component of the UFO, which is now unresolved in a blend of multiple absorption lines probably from Ne and Fe ii) constrain with unprecedented accuracy the properties of the high energy UFO components by resolving individual Fe XXV/XXVI absorption lines iii) shed light on the launching mechanism if, as expected, the detailed shape of the absorption features are different for magnetic, radiation, and thermal driving \citep{Fukumura22}.

\begin{acknowledgements} 
We thank the anonymous referee for their valuable feedback, which significantly enhanced the quality of this manuscript.
GL, GAM, and MB acknowledge
support and funding from Accordo Attuativo ASI-INAF n. 2017-14-H.0. 
GL, EB, MC and MD acknowledge support from PRIN MIUR 2017 Black Hole Winds and the Baryon Life Cycle of Galaxies, 2017PH3WAT. 
EB acknowledges the support of the INAF Large Grant 2022  “The metal circle: a new sharp view of the baryon cycle up to Cosmic Dawn with the latest generation IFU facilities”.
FT acknowledges funding from the European Union - Next Generation EU, PRIN/MUR 2022 (2022K9N5B4). 
AL acknowledges support from the HORIZON-2020 grant “Integrated Activities for the High Energy Astrophysics Domain" (AHEAD-2020), G.A. 871158).
JR and VB acknowledge support through NASA grant 80NSSC22K0474.
MG acknowledges partial support by NASA HST GO- 15890.020/023-A, and the \textit{BlackHoleWeather} program.
RS acknowledges support from the agreement ASI-INAF eXTP Fase B - 2020-3-HH.1-2021 and the INAF-PRIN grant “A Systematic Study of the largest reservoir of baryons and metals in the Universe: the circumgalactic medium of galaxies” (No. 1.05.01.85.10).
MD gratefully acknowledges INAF funding through the "Ricerca Fondamentale 2022”  program (mini-grant 1.05.12.04.04).
MP acknowledges grant PID2021-127718NB-I00 funded by the Spanish Ministry of Science and Innovation/State Agency of Research (MICIN/AEI/ 10.13039/501100011033).
\end{acknowledgements}

\thispagestyle{empty}
\bibliographystyle{aa}
\bibliography{references}

\begin{appendix}

\section{1-4 keV Spectral Scan}
\label{apsec:scan}

In \autoref{figapp:scan_med}, we report the results of the line scan performed in the intermediate energy range (1.5-4 keV) where MgII, Si XIV, and SXVI \lya absorption lines produced by UFOs are sometimes observed, even in CCD-resolution spectra \citep[e.g.,][]{Pounds03,Parker18iras}. 
The continuum is the same as described in Sec. \ref{subsec:continuum} while the details of the scan are the same as described in Sec. \ref{subsec:spec_scan}.
Indeed, three narrow residuals can be seen at energies roughly corresponding to the expected energies of such lines for an outflow velocity of $\vout=0.27c$ (dashed lines in \autoref{figapp:scan_med}). The significance of these features is, however, low and up to 99\% confidence level only for the S XVI \lya line. 
The \dc values from the fit with a Gaussian line for each of these features are 4.27, 3.17, and 8.44, respectively.
They contribute, nonetheless, to the total \dc when the \xstar or \dwind tables are added to the continuum. 
As a consequence, including these complex absorption templates into the fit results in a significantly increased statistical significance compared to the fit with the single Gaussian lines (see Sec.~\ref{subsec:xstar}).

\begin{figure}
\centering
\includegraphics[width=0.46\textwidth]{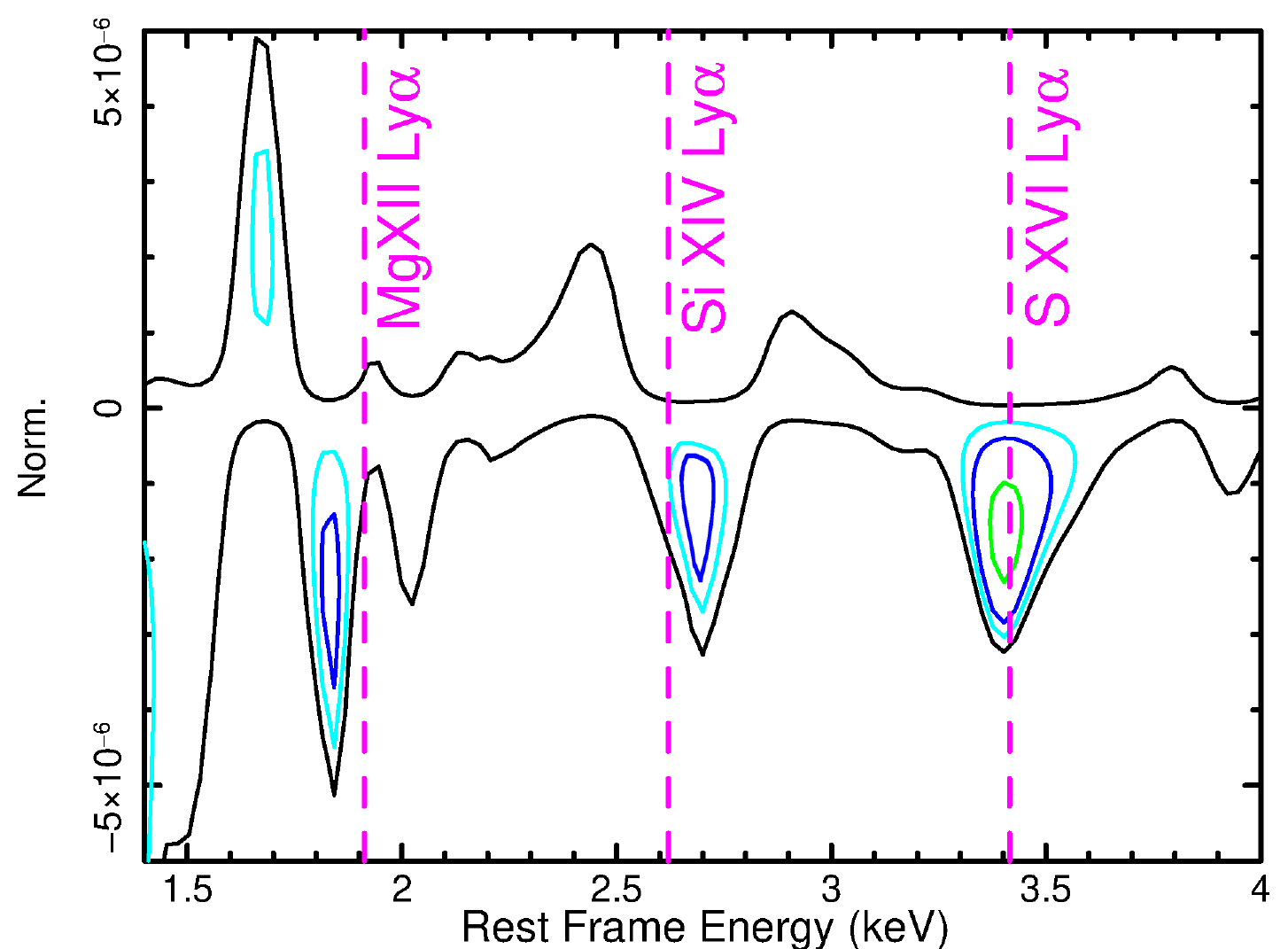}
\caption{Confidence contours showing deviations from the continuum model described in Sec.~\ref{subsec:continuum} in the intermediate band. Contours as in \autoref{fig:zoomSH}.}
\label{figapp:scan_med}
\end{figure}

\section{MonteCarlo simulations}
\label{apsec:montecarlo}
For our null hypothesis baseline model, we used the $0.6-24$ keV rest frame best-fit continuum model described in Sec.~\ref{subsec:continuum}. We simulated 1000 sets of \xmm and \nustar spectra with the same response matrices, exposure times, and background levels as the read data. These spectra were binned following the same \kb binning used for the observed spectra. We then fitted the continuum model to each set of simulated spectra, leaving the same free parameters described in Sec.~\ref{subsec:continuum}. Then, we simulated the spectra again, starting from the new best-fit values to accommodate for random continuum variations. We fitted these newly generated spectra again to find the new continuum best-ft (see \citealt{Matzeu23} for a similar procedure).

We perform a second fit on the same set of simulated spectra, adding a narrow Gaussian profile, with line width fixed to zero and normalization set to zero but free to fluctuate between negative and positive values 
to probe both absorption and emission features. The rest-frame Gaussian line energy is then stepped with the \textsc{steppar} command in \xspec, between 
0.5 and 2 keV in $\Delta E=15\ev$ for the soft band features (Em$_1$ and Abs$_1$) and between 5 and 10 keV in $\Delta E=25\ev$ for the hard band features (Em$_2,_3$ and Abs$_2,_3$) to find the global $\cstat$ minimum for each set of spectra in the respective energy ranges. 
We verified that extending the hard band range to, e.g., 12 keV to take into account that Abs$_3$ is observed above 10 keV would not change the $\mathcal{MC}$ results, as the strongest random fluctuations always occur where the count statistic is larger, i.e., toward the low energy boundary, between $\sim5$ and $\sim7$ keV. Interestingly, this also implies that by performing the search on the full 4-10 keV band - as usually done in the literature - we tend to underestimate the significance of the UFO features, as by definition, they appear above $\sim7.5$ keV (see also \citealt{Nardini19}).

This process maps the \dc variations due to random fluctuations in the simulated spectra. The $\mathcal{MC}$ statistical significance (\pmc) of the emission/absorption line detections can be then computed as \pmc$=1-\left(\frac{N}{S}\right)$ where N is the number of simulated spectra showing a random fluctuation larger than the one observed in the data and S the number of simulations.

None of the realizations in the soft band show \dc$>37.96$ and therefore Em$_1$ has \pmc$>99.9\%$, while 2 realizations have \dc$>18.56$,  indicating \pmc$=99.8\%$ for Abs$_1$.
None of the realizations in the hard band show \dc$>22.28$ and therefore Em$_2,_3$ and Abs$_2$ all have \pmc$>99.9\%$ while 2 exceed \dc$=16.59$ and therefore Abs$_3$ has \pmc$=99.8\%$. Even the less prominent absorption features, Abs$_1$ in the soft band and Abs$_3$ at 11 keV, are significant at $>3\sigma$ from \mc simulations.

Having performed 1000 sets of simulations for the soft and hard band, respectively, the accuracy of \pmc is limited to the level of 1/1000. However, significantly increasing the number of trials to increase the accuracy of \pmc is computationally prohibitive due to the quite complex baseline model, the simultaneous fit of five different spectra, and the need to sample two different energy ranges.

\begin{figure}
\centering
\includegraphics[width=0.45\textwidth]{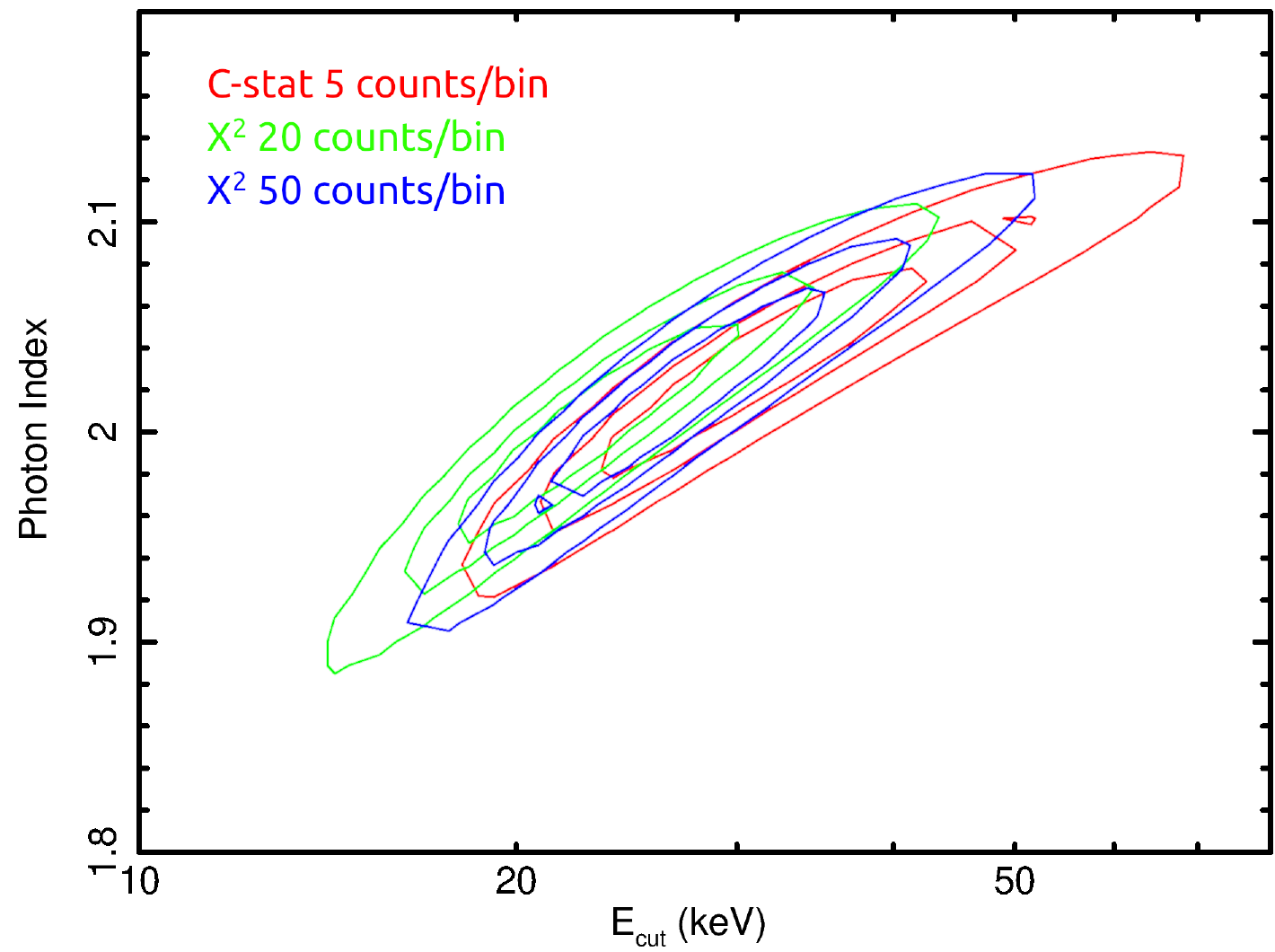}
\caption{Contour plots of high energy cutoff vs. photon index for different binning schemes and statistics adopted. In red, the results from a fit performed with $\cstat$ on spectra binned to 5 counts per bin, and in blue (green) from a fit performed with $\chisq$ on spectra binned to 20 (50) counts per bin.}
\label{figapp:ecut}
\end{figure}

\section{Tests on $\ecut$ measurements}
\label{apsec:ecut}

For the spectral fits performed in this work, we used the \kb binning that optimizes the bin size based on the spectral resolution and the source counts available per resolution element, with the main goal of characterizing the emission and absorption features observed. However, to our knowledge, this binning scheme has never been tested for continuum parameters accuracy, with the recent exception of \citet{Zappacosta23} where, however, it was tested in a completely different regime: extreme low counts statistic and simple power-law models.

We, therefore, tested the impact of different fit statistics and spectral binning on the continuum parameters $\Gamma$ and $\ecut$ with particular emphasis on the extremely low high energy cut-off found in \autoref{sec:coronal}.
We computed the $\Gamma$ vs. $\ecut$ contours for the $\cstat$ fit on \xmm, and \nustar spectra binned to 5 counts per bin, and for $\chi^2$ fits on spectra binned to 20 and 50 counts per bin. As can be seen in \autoref{figapp:ecut}, the results are stable, independently of the binning scheme and statistics adopted for the fit. We also tested the impact of including the energy ranges 20-30 keV and 20-50 keV, where the \nustar spectra are background-dominated but where some source signal is still present. Once again, the results are stable, with $\ecut$ in the range $25-30\pm10$ as for the fit limited to 20 keV observer-frame.

\end{appendix}

\end{document}